\documentclass[lettersize,journal]{IEEEtran}
\usepackage[switch]{lineno}

\usepackage[hidelinks=false]{hyperref}
\hypersetup{
    colorlinks=true,
    linkcolor=blue,
    citecolor=blue,
    urlcolor=blue
}

\usepackage{amsmath,amsfonts}
\usepackage{array}
\usepackage{textcomp}
\usepackage{stfloats}
\usepackage{url}
\usepackage{verbatim}
\usepackage{graphicx}
\usepackage{cite}
\usepackage{tikz}
\newcommand{\stepnum}[1]{%
  \tikz[baseline=(X.base)] \node (X) [draw, circle, inner sep=0.4pt] {\footnotesize #1};%
}

\usepackage{cleveref}

\begin{document}

\title{DT-RAID: A Software-Defined Tiered RAID Architecture for Heterogeneous SSDs}

\linespread{0.89}


\author{
\centering
\begin{tabular}{ccc}
\textbf{Kun-Chi Chiang} & \textbf{Radu Stoica} & \textbf{Animesh Trivedi} \\
Macronix Inc. & IBM Research Zurich & IBM Research Zurich \\
kunchichiang@mxic.com.tw & rst@zurich.ibm.com &Animesh.Trivedi@ibm.com\\[1.0em]

\textbf{Chun-Lien Su} & \textbf{Liang-Chi Chen} & \textbf{Roman Pletka} \\
Macronix Inc. & National Taiwan University & IBM Research Zurich \\
juliensu@mxic.com.tw & d12922012@csie.ntu.edu.tw & rap@zurich.ibm.com
\\[1.0em]

\multicolumn{3}{c}{
\begin{tabular}{cc}
\textbf{Wei-Kuan Shih} & \textbf{Chien-Chung Ho} \\
National Tsing Hua University & National Cheng Kung University \\ 
wshih@cs.nthu.edu.tw & ccho@gs.ncku.edu.tw 
\end{tabular}
}
\end{tabular}
}



\maketitle

\begin{abstract}
The rapid proliferation of cloud and AI-driven workloads has led to increasingly complex requirements for modern storage subsystems. To meet these demands, SSD controller architectures have evolved into a fragmented landscape, offering tiers of drive types optimized for endurance, performance, or capacity. More recently, SSDs have begun to differentiate regions within the same device, enabling intra-drive heterogeneity. However, integrating such heterogeneity into the existing storage stack with minimal disruption remains challenging.

In this paper, we argue that storage middleware, such as RAID, is an effective control layer to address these integration challenges. We present DT-RAID, an intra-drive heterogeneity-aware RAID architecture designed for emerging SSDs. DT-RAID monitors stripe-level I/O access patterns and makes online placement decisions without requiring application modifications. It employs a lightweight heat-tracking mechanism to dynamically place frequently accessed (hot) stripes onto the higher-performance, higher-endurance tier. Using simulations based on SNIA MSR enterprise I/O traces, we demonstrate that DT-RAID improves modeled I/O performance by up to $6.8\times$ under greater tier asymmetry and extends normalized lifespan by up to $20.9\times$ compared to uniform RAID deployments.

\end{abstract}

\section{Introduction}
\IEEEPARstart{M}{odern} data-intensive applications demand higher I/O performance and lower storage costs, while becoming more diverse, evolving more rapidly, and having requirements that are increasingly harder to predict~\cite{li2020cloudblock,10.1145/3297663.3310302,ZhichaoCao2020RocksDBworkloads,berg2020cachelib,tang2023cloudonomics}. For example, in a cloud environment, the same infrastructure can simultaneously support data lakehouses with long-running batch and streaming pipelines, database services or key–value stores that process high-frequency transactions, and AI training and inference workloads~\cite{Dong2024CloudNativeDatabases,Xue2024LakehouseScale,Hu2024CharLLMinDatacenter}. 

NAND-based Solid-State Drives (SSDs) have become the dominant storage medium for modern performance-critical applications. To serve increasingly diverse workload requirements, the SSD ecosystem has evolved toward specialized device offerings, including low-latency drives~\cite{KIOXIAFL6Datasheet}, balanced enterprise SSDs~\cite{SolidigmD7P5810Brief}, and high-density cold-storage SSDs\cite{MicronION6500}. Beyond such device-level specialization, recent SSD designs increasingly integrate heterogeneous flash technologies within a single device. This intra-drive heterogeneity creates new opportunities for buffering, data placement, and internal resource optimization~\cite{Radu2019HybridFlashCTRL,Zgang2019SPASSD,Wu2024FIRMTree}.

Although such heterogeneity is still largely hidden behind the conventional block interface today, emerging host-managed storage interfaces make it increasingly plausible for the host to observe and exploit intra-drive heterogeneity~\cite{2025MixedModeSSD,2025scalaafa-journal,2025KVSSD_park,kang2014multistreamssd,NVMeBaseSpec22}. These developments extend the traditional block-device abstraction and open the possibility of host-driven data placement and dynamic reconfiguration of SSD resources~\cite{BjorlingBBD13}. One example is Kioxia's mixed-mode SSD proposal, which exposes multiple capacity- and performance-optimized tiers that can be reconfigured according to workload demands~\cite{2025MixedModeSSD}. 
However, this flexibility is fundamentally constrained by the fixed amount of flash resources within each SSD. Allocating more resources to performance modes inevitably reduces usable capacity, turning heterogeneous flash management into a system-level trade-off rather than a free optimization. This, in turn, raises the question of where such management decisions should be made.

A natural way to exploit intra-drive heterogeneity is to push placement decisions to applications or into SSD firmware, but both options are problematic. Application-level adaptation requires modifying each application individually, creating high engineering costs and tight coupling to hardware-specific features. In contrast, SSD-internal tiering lacks application- and system-level visibility, making it difficult to infer true data importance across concurrent workloads, while also being constrained by limited controller resources and often yielding unstable performance~\cite{Yoo2020Reinforcement, Boukhobza2025HostSideFlash}.

To strike a balance between these two extremes, we focus on the RAID middleware, which preserves the standard block interface while remaining close enough to the devices to enable placement control~\cite{2022scalaraid,Shu2023dRAID,2025scalaafa-journal}. We extend the RAID layer with heterogeneity-aware placement to support intra-drive heterogeneity. Incorporating such placement into the RAID layer provides four benefits: 
(1) no application or filesystem changes are required, since the RAID layer preserves the existing block interface; 
(2) the RAID layer has a global cross-device view of access patterns, enabling stripe-level hotness tracking without fragmented placement decisions;
(3) the RAID layer already reasons about parity-induced I/O amplification and can therefore distinguish data placement from parity placement; and 
(4) the design introduces minimal SSD dependencies. 
Taken together, these properties make the RAID layer a practical control point for multi-tier SSDs and raise the question of stripe placement under parity-based update semantics.

In this paper, we study how intra-drive heterogeneity can be exploited at the RAID layer. In parity-based RAID, maintaining parity consistency introduces additional I/O operations beyond user requests, and this parity overhead is workload-dependent, potentially accelerating wear under skewed workloads. To capture this effect, we develop a lightweight heat-tracking mechanism that monitors stripe-level update intensity and guides the differentiated placement of parity and data across heterogeneous flash tiers.

We then propose DT-RAID (Dual-Tier RAID), which manages data placement across performance- and capacity-optimized tiers within an SSD array. We focus on the dual-tier setting because it captures the essential performance-capacity trade-off. Since commodity SSDs do not yet expose such heterogeneity directly to the host, we evaluate DT-RAID using a trace-driven RAID simulation platform. Using enterprise I/O traces from the SNIA IOTTA repository~\cite{msr_iotta}, we show that DT-RAID improves modeled I/O performance by up to $6.8\times$ relative to the uniform-flash baseline and extends normalized lifespan by up to $20.9\times$ when tier endurance asymmetry is sufficiently large, with bounded capacity reduction.

The contributions of this paper are as follows:
\begin{itemize}
    \item We identify the opportunity for RAID-layer data placement on emerging SSDs that expose intra-drive heterogeneous tiers and motivate stripe-aware management as a practical system-level approach.
    \item We design DT-RAID, including a lightweight heat-tracking mechanism and dual-tier placement policies for SSD arrays with host-visible heterogeneous flash tiers. 
    \item We implement a trace-driven RAID simulator and provide a detailed evaluation of DT-RAID across enterprise workloads, quantifying its benefits in modeled I/O performance, normalized lifespan, and capacity reduction.
\end{itemize}

The paper is organized as follows. 
In \Cref{sec:background}, we present the background of this work, covering flash operating modes, RAID architecture and its management overheads, and workload skew characteristics.
In \Cref{sec:relatedwork}, we review the related work. 
In \Cref{sec:design_principles}, we highlight the DT-RAID design principles. 
In \Cref{sec:dt_raid_design}, we present the DT-RAID design concepts.
In \Cref{sec:simulator_arch}, we describe the simulator architecture.
In \Cref{sec:evaluation}, we evaluate and quantify the potential multi-dimensional benefits of DT-RAID.
Finally,~\Cref{sec:conclusion} concludes the paper.

\section{Background} \label{sec:background}

\subsection{NAND Flash Operating Modes}
\label{sec:flash_operating_mode}

\begin{table}[t]
\footnotesize
\caption{Representative latency and endurance characteristics of NAND flash memory~\cite{ren2026deviceleveloptimizationtechniquessolidstate,Kim2025REO,Yoo2020Reinforcement}.}
\label{tab:tier-specs}
\setlength{\tabcolsep}{3pt}
\begin{tabular}{l c c c c}
\hline
Parameter              & SLC           & pSLC          & TLC           & QLC \\ \hline
Read latency (16kB)     & 20--25\,$\mu$s  & 25--30\,$\mu$s  & 45--170\,$\mu$s & 120--200\,$\mu$s \\
Program latency (16kB)  & 50--220\,$\mu$s & 50--220\,$\mu$s & 0.4--2\,ms      & 2.0--3.0\,ms \\
Block erase latency    & 1.5--5\,ms      & 1.5--5\,ms      & 3.5--15\,ms      & 15--20\,ms\\
Endurance (P/E Cycles) & 60--100\,K      & 10--50\,K       & 3--6\,K        & 0.5--3\,K \\
\hline
\end{tabular}
\end{table}
Modern 3D NAND flash supports multiple operating modes that differ in bit density per cell, leading to distinct trade-offs in performance, endurance, and capacity~\cite{ren2026deviceleveloptimizationtechniquessolidstate}. SLC and pseudo-SLC (pSLC) modes enable fast, durable operation by storing one bit per cell, with pSLC dynamically converting higher-density multi-bit flash into single-bit mode to deliver a high-performance tier~\cite{Yoo2020Reinforcement}. In contrast, TLC and QLC flash offer higher density at the cost of higher latency and lower endurance. Among these modes, TLC is widely adopted in enterprise SSDs because it provides a practical balance among capacity, performance, and device endurance.

\Cref{tab:tier-specs} summarizes representative latency and endurance ranges across these operating modes. As the number of bits per cell increases, read latency rises because more threshold-voltage levels must be distinguished during sensing, and program latency increases because tighter voltage control and additional verification steps are required. Endurance also decreases with higher bit density, since narrower voltage margins make cells more vulnerable to wear and retention-related errors. As a result, SLC-like tiers offer lower latency and higher endurance, whereas TLC- or QLC-like tiers provide higher capacity at lower cost\cite{Cai2017ErrorCharacters,ren2026deviceleveloptimizationtechniquessolidstate}.


These asymmetries are fundamental to SSD designs with \textit{intra-drive heterogeneity}. They can be further amplified by mode-specific operating methods and device-level optimizations, including shallow erase or shallow program in pSLC tiers, that reduce wear and improve write efficiency~\cite{Cho2024AERO,Ren2024NearFree}. Thus, flash heterogeneity arises not only from intrinsic cell density, but also from how each tier is provisioned and operated, making heterogeneous flash tiers meaningful resources for higher-layer performance and endurance optimization.

\subsection{RAID Basics and Parity Update Overheads}
\label{sec:background_raid_manage_overhead}
A Redundant Array of Independent Disks (RAID) is a fundamental storage architecture that aggregates multiple physical drives into a logical unit. A RAID system improves performance through parallelism, increases storage capacity, and enhances availability through redundancy. Different RAID schemes represent distinct trade-offs among usable capacity, bandwidth, and fault tolerance. 

In this paper, we focus on parity-based RAID schemes, RAID-5 and RAID-6, which are some of the most widely deployed configurations in practice as they tolerate one or two drive failures and have substantially lower storage overhead than mirroring. 
RAID schemes typically operate on fixed-size units (chunks or stripe units): a stripe consists of one chunk from each of the $N$ member drives. 

Parity-based RAID schemes introduce additional I/O amplification due to \textit{parity updates}. For RAID-5, each stripe contains $N-1$ data chunks and one parity chunk, where parity encodes the XOR of the data chunks and rotates across drives for load balancing. To guarantee data recoverability under a drive failure, the parity and user data chunks must remain consistent, requiring additional I/O operations. 
RAID-6 extends RAID-5 by maintaining two independent parity chunks, typically implemented using Reed-Solomon-style coding, and therefore incurs higher parity-update overhead than RAID-5.

RAID I/O amplification depends heavily on the size of the write operations relative to the stripe size. For example, when a write updates all $N\!-\!1$ data chunks in a stripe, a RAID scheme can update the entire stripe along with the corresponding parity chunks. This scenario is referred to as a \emph{full-stripe write} (FSW) and incurs the minimum parity-update overhead. In contrast, when a write updates fewer than $N\!-\!1$ data chunks, the RAID layer must preserve parity consistency via \emph{partial-stripe} updates, typically using \emph{read--modify--write} (RMW) or \emph{read--reconstruct--write} (RCW). RMW first reads the old data to be overwritten and the old parity, computes a parity delta, and then writes the updated data and parity. RCW instead reads the remaining unchanged data chunks to reconstruct the new parity before issuing the writes. 
\begin{figure}
    \centering
    \includegraphics[width=1\linewidth]{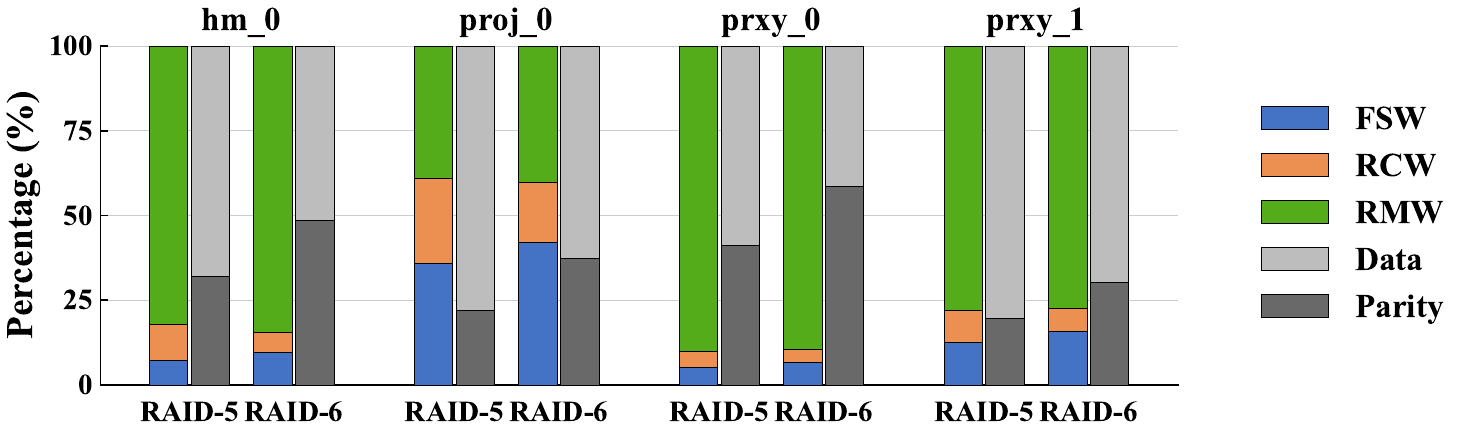}
    \caption{RAID update-strategy breakdown (FSW/RCW/RMW) and overall data/parity operation breakdown for four write-intensive traces under 8-disk RAID-5 and RAID-6 with 4 KB chunks.}
    \label{fig:bak_io_data_in_raid}
\end{figure}

\Cref{fig:bak_io_data_in_raid} shows the breakdown of I/O activity into data and parity operations for the four write-intensive traces among the 20 largest traces in SNIA IOTTA~\cite{msr_iotta} (Table~\ref{tab:datasets} summarizes these traces, and Section~\ref{sec:exp-method} describes the experimental methodology). Each trace is represented by two adjacent bars. The first bar breaks down parity update activity by update strategy. The second bar shows the overall proportion of data and parity operations in the corresponding trace. We observe that traces with higher fractions of RMW and RCW, such as \texttt{hm\_0} and \texttt{prxy\_0}, exhibit a larger share of parity-induced I/O activity. For example, \texttt{prxy\_0} is dominated by partial-stripe updates, with 90.2\% RMW and 4.6\% RCW, and correspondingly shows 41.1\% parity-related operations. In contrast, \texttt{proj\_0} has 35.7\% parity updates handled by FSW, and its parity-related activity is lower at 22.1\%. Since parity updates introduce additional backend I/O beyond user data updates, the frequency and locality of parity updates can materially affect RAID efficiency under write-heavy workloads.

\subsection{I/O Workload Skew}
\label{sec:background_wkld_skew}
A substantial body of prior work has collected I/O traces from production storage systems and analyzed their statistical properties. Examples include studies on enterprise storage workloads~\cite{msr_iotta, yang2016write, ZhichaoCao2020RocksDBworkloads}, cloud storage systems~\cite{berg2020cachelib, li2020cloudblock, wu2025skewness}, and workloads targeting SSD-based infrastructures~\cite{yadgar2021ssd, maneas2022operational}. A consistent observation across these independent studies is that real-world I/O accesses are inherently skewed: a small fraction of logical blocks accounts for a disproportionately large share of read and write activity.

\begin{figure}[tb]
  \centering
  \begin{minipage}{\linewidth}
    \centering
    \includegraphics[width=0.8\linewidth]{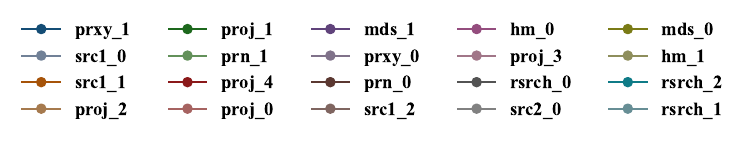}\\
    \begin{minipage}[t]{0.5\linewidth}
      \centering
      \includegraphics[width=\linewidth]{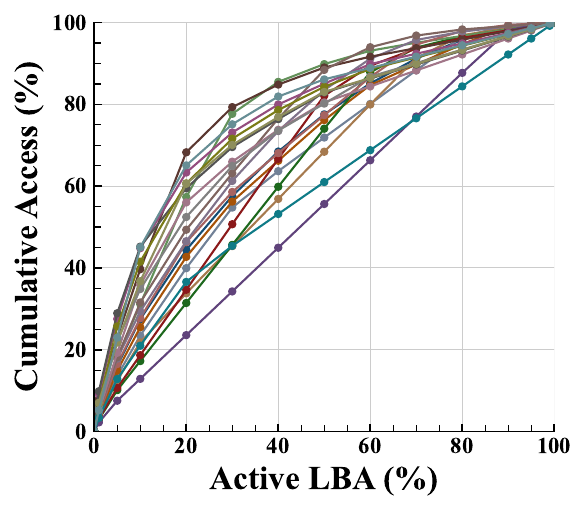}\\
      {\small (a) Active LBA CDF.}
      
    \end{minipage}\hfill
    \begin{minipage}[t]{0.5\linewidth}
      \centering
      \includegraphics[width=\linewidth]{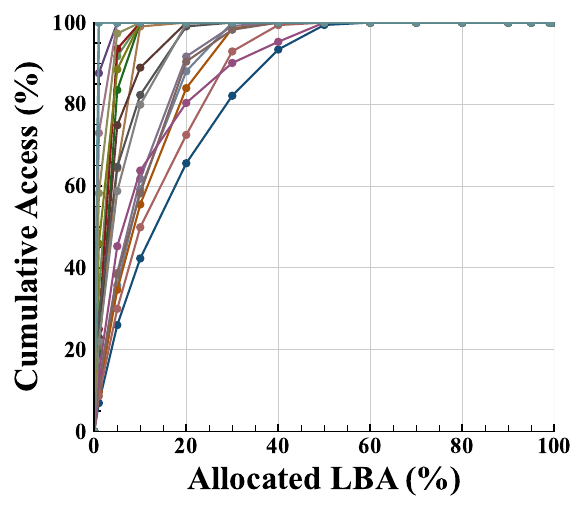}\\
      {\small (b) Allocated LBA CDF.}
    \end{minipage}
  \end{minipage}
    
  \caption{Cumulative access-frequency distributions of the 20 largest SNIA traces.}
  \label{fig:snia-workload-skew}
\end{figure}

To quantify the degree of skew present in realistic workloads, we analyze the MSR SNIA I/O trace suite~\cite{msr_iotta}. Specifically, we select the top 20 largest traces from the dataset and compute the cumulative access frequency distribution. The resulting distribution is illustrated in \Cref{fig:snia-workload-skew}.

Storage systems provision the full logical address space, whereas a time-bounded workload trace typically accesses only a small fraction of it. To distinguish these views, we analyze workload skew over both the \textit{active} and \textit{allocated} LBA spaces. In \Cref{fig:snia-workload-skew}(a), the x-axis denotes the fraction of LBAs in the active LBA set, sorted by access frequency, and the y-axis denotes the cumulative fraction of accesses contributed by those LBAs. In \Cref{fig:snia-workload-skew}(b), the x-axis instead denotes the fraction of the allocated LBA space up to the maximum observed LBA, including addresses that were never accessed, while the y-axis again shows the access fraction.

Skew is already visible in the active-LBA view but appears less pronounced because unused addresses are excluded. When the full allocated LBA space is considered, the skew becomes much stronger across nearly all traces: a small fraction of the logical space accounts for most I/O activity, while most provisioned space remains cold. These results confirm that skewed access distributions are pervasive in real-world workloads.




\subsection{Implications of Skewed Workloads for Parity-Based RAID}
\label{sec:implic_work_skew}
In parity-based RAID systems, parity consistency is maintained at stripe granularity. As a result, repeated updates to a small set of logical addresses tend to concentrate activity on a subset of RAID stripes. Because each update requires additional parity updates within the same stripe, logical write skew propagates into skewed RAID-side activity across stripes.

This effect is particularly important in heterogeneous flash environments, where different storage tiers provide different performance and endurance characteristics. Under such conditions, stripe-level access skew can translate into uneven flash activity across tiers, amplifying tier-level imbalance and long-term wear stress. Because parity-based RAID enforces updates at stripe granularity, the cost of each update depends on the tier composition of the corresponding stripe.
These observations motivate our work on RAID-level placement mechanisms that explicitly account for stripe-level update skew across heterogeneous flash tiers.
\section{Related Work}
\label{sec:relatedwork}
We organize prior work into three categories: homogeneous software RAID, RAID with \textit{inter-drive} heterogeneity, and RAID with \textit{intra-drive} heterogeneity. This classification highlights how existing studies differ in whether they assume uniform SSD arrays, exploit heterogeneity across drives, or leverage performance and endurance tiers within a single SSD.

\textbf{Homogeneous software RAID.}
A large body of prior work studies software RAID over homogeneous SSD arrays. These designs primarily focus on removing software bottlenecks in the I/O stack and improving scalability on modern NVMe devices. Representative examples include ScalaRAID and STRAID~\cite{2022scalaraid,2024straid-journal}, which show that a well-designed software RAID stack can match or even outperform traditional hardware RAID controllers. However, these systems largely assume that all SSDs expose similar latency, endurance, and capacity characteristics. They improve RAID efficiency under uniform drive configuration assumptions, but do not address placement when flash heterogeneity is visible to the host.

\textbf{RAID with inter-drive heterogeneity.}
As SSDs diversify across performance, endurance, and capacity tiers such as SLC, TLC, and QLC, recent studies have explored RAID designs that exploit \emph{inter-drive heterogeneity}, where different drives in the array are built with different flash technologies. HybRAID, for example, separates frequently updated data from colder data by placing hot data on an SLC-based RAID-1 tier and cold data on a TLC-based RAID-5 tier, thereby reducing parity-update overheads on endurance-limited devices~\cite{2024hybraid}. Asymmetric RAID further relaxes the homogeneity assumption by allowing SSDs with different performance and capacity characteristics to coexist within a single RAID array~\cite{2024asymetricraid}. These designs show that heterogeneity can be exploited above a uniform-array abstraction, but they primarily treat it as a \emph{device-level composition problem}: deciding which class of drives should serve which RAID role or tier. In contrast, we treat heterogeneity as a \emph{RAID-level placement problem}, asking how host-visible flash tiers should be used as explicit placement targets for stripe components under parity-coupled update costs.

\textbf{RAID with intra-drive heterogeneity.}
More recently, RAID designs built on SSDs with host-visible heterogeneous flash tiers have begun to explore \emph{intra-drive heterogeneity}, where a single SSD exposes a small high-performance tier alongside a larger capacity-oriented tier. Prior work has shown that such tiers can be exposed to the host and selectively used for performance- or endurance-critical data~\cite{2025MixedModeSSD}. Building on this capability, ScalaAFA uses an SLC partition as a staging area for parity updates and later materializes parity with device-side support~\cite{2025scalaafa-journal}. This line of work demonstrates the practicality of host-visible heterogeneous flash, but it still treats the faster tier mainly as a \emph{temporary staging space} rather than a \emph{persistent placement tier}. In contrast, we treat heterogeneous flash tiers as persistent RAID-level placement targets and use them to directly place stripe components within the RAID layer.

\textbf{Summary}. Prior work exploits flash heterogeneity in RAID mainly in two ways: composing arrays from heterogeneous drives or using faster flash tiers as temporary staging space. Our work takes a different view. We propose to treat host-visible heterogeneous flash as a substrate for \emph{persistent stripe-level placement} within the RAID layer. This perspective allows us to study how parity semantics, workload skew, and tier asymmetry jointly shape placement decisions, performance, and long-term wear.
\section{DT-RAID Design Principles}
\label{sec:design_principles}

\subsection{Intra-Drive Heterogeneity Model}
As discussed in \Cref{sec:flash_operating_mode}, modern NAND flash supports multiple operating modes with different latency and endurance characteristics. DT-RAID captures this heterogeneity using a dual-tier abstraction, where each SSD is modeled as exposing two host-visible tiers with distinct performance and endurance properties. This preserves the essential performance-capacity trade-off while keeping the placement problem tractable. Such tiers can be realized today with a mix of SLC (performance optimized), TLC or QLC (capacity optimized) NAND technologies. Extending DT-RAID to more than two tiers is beyond the scope of this work.
All placement decisions are made at the RAID layer, which preserves the block interface while exposing redundancy-aware control.


\subsection{Cost-Aware Placement Objective}

As discussed in \Cref{sec:implic_work_skew}, stripe placement in heterogeneous flash RAID is shaped by three factors: flash cost asymmetry, parity-induced structural overhead, and workload-driven skew. In heterogeneous flash, reads and writes incur different latency and endurance costs across tiers. In parity-based RAID, partial-stripe updates introduce parity-related overhead beyond user I/O. Under skewed workloads, this overhead tends to concentrate on a subset of stripes. Accordingly, DT-RAID aims to place stripes based not only on access frequency, but also on their effective cost under these three factors.

To make such decisions, DT-RAID considers several properties that determine the effective cost of serving a stripe. First, access recency and frequency remain important because they indicate which stripes are repeatedly touched and thus are more likely to benefit from faster tiers under skewed workloads. Second, reads and writes have asymmetric costs in NAND flash. As discussed in \Cref{sec:flash_operating_mode}, reads are typically faster and less damaging than program and erase operations. Consequently, write-dominant stripes should be treated as more cost-sensitive than read-dominant stripes in tier placement. Finally, placement must account for amplification across the storage stack. At the RAID layer, partial-stripe updates can trigger parity-related overhead, while at the device layer, internal management can further increase the effective write cost. DT-RAID therefore seeks to minimize the \textit{tier-weighted stripe service cost}, rather than relying solely on access frequency, by jointly considering access intensity and RAID-induced flash costs.

\subsection{Stripe Granularity as the Placement Unit}
We adopt stripe-level placement for four reasons. First, the stripe is the natural unit of RAID management because it captures cross-device locality and matches the granularity of parity maintenance, caching and coalescing, journaling, and rebuild operations. Second, stripe-level accounting keeps metadata manageable. Under our baseline (8-disk RAID-5, 4\,KiB chunks, 8\,TiB per device), the array has about 2.15 billion stripes; two 8-bit read/write counters per stripe require about 4\,GiB of DRAM, and larger chunk sizes further reduce this cost. We also maintain a 1-bit per-stripe bitmap to mark hot stripes (about 0.25\,GiB), which simplifies hot-stripe lookup.
Third, stripe-level placement is fine-grained enough for practical tier separation while avoiding the complexity of chunk-level tracking.
Finally, stripe size is already a familiar and tunable RAID parameter, making stripe-level placement compatible with existing storage-system configuration practices.

\subsection{Design Scope and Abstraction}
\label{sec:modeling_scope_and_limit}
This work studies RAID-layer stripe management for SSDs with intra-drive heterogeneity. In particular, we study how stripe-level placement decisions shape SSD-level I/O behavior under enterprise workloads and influence the performance-endurance trade-off across heterogeneous flash tiers.

Accordingly, rather than simulating the full SSD stack, the model intentionally abstracts away device-internal mechanisms such as garbage collection, block allocation, wear leveling, and background maintenance. Our goal is to isolate the causal impact of RAID-layer stripe management. Incorporating full device-internal behavior would introduce additional nondeterministic effects that could obscure controlled comparisons across placement policies.


As a result, our analysis captures only RAID-layer activity and does not include transient interference or secondary write amplification caused by SSD-internal management. Although latency or lifetime values may differ on real devices, the trends in parity-update-induced amplification, stripe access skew, and tier-dependent activity imbalance remain informative. In practice, internal SSD mechanisms such as garbage collection may further amplify skewed updates~\cite{Boukhobza2025HostSideFlash,Kim2025D2FS}, suggesting that the skew-related trends in our analysis may become more pronounced when such effects are considered.

Finally, we assume host-visible flash tiers with configurable performance and endurance budgets. Although such configurations are not yet universally available in commodity SSDs, they are increasingly discussed in research prototypes and emerging products. Our study, therefore, focuses on the implications of host-visible tier asymmetry for RAID-layer design, regardless of any specific commercial implementation.
\section{Dual-Tier RAID (DT-RAID) Design}
\label{sec:dt_raid_design}
\subsection{Design Space: Static Versus Dynamic Placement}
\label{sec:design_space}
The preceding analysis shows that tier-aware RAID placement must address two distinct sources of cost under host-visible flash heterogeneity. The first is \emph{structural}: parity-based RAID necessarily generates parity writes to preserve redundancy consistency, even under uniform workloads. The second is \emph{workload-driven}: skewed updates that dynamically target a small subset of stripes, causing flash activity and parity-update overhead to become concentrated over time.

These two cost sources motivate two complementary placement strategies. \emph{Parity-driven static} placement targets the structural parity-write cost of parity-based RAID: because each user write updates both data and parity, parity-related writes persist regardless of workload skew. Placing parity on a higher-endurance tier therefore reduces unnecessary wear on the lower-endurance tier without changing RAID geometry or redundancy semantics. \emph{Workload-driven dynamic} placement targets workload-induced skew: when a small subset of stripes receives disproportionate updates, adaptive stripe-level migration can steer hot stripes to the performance tier, reducing access latency and mitigating uneven wear.

Accordingly, DT-RAID provides two tier-placement policies. \textit{Static RAID} (DT-S) persistently places parity on the higher-endurance tier and isolates the workload-independent benefit of parity placement. \textit{Dynamic RAID} (DT-D) instead uses stripe heat to decide whether an entire stripe should reside in the performance tier or the capacity tier. Both policies preserve RAID geometry, parity rotation, and redundancy semantics; they differ only in how stripe components are mapped to flash tiers.

\begin{figure}[t]
    \centering
    \includegraphics[width=0.9\linewidth]{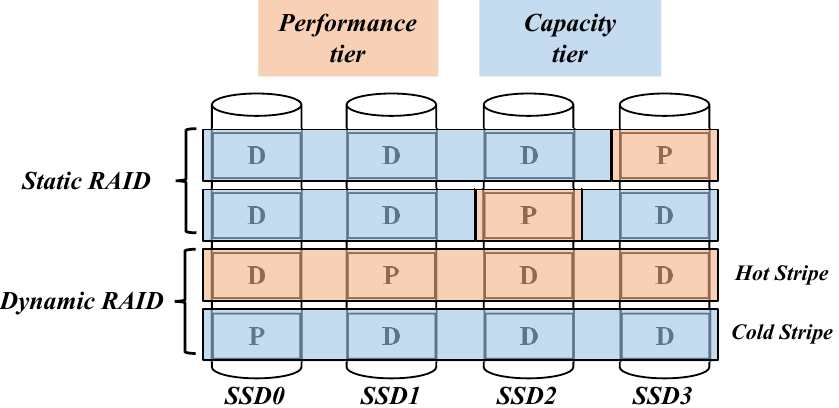}
  \caption{RAID-5 stripe-level placement in DT-RAID. The top two rows show \textit{Static RAID}; the bottom two rows show \textit{Dynamic RAID}. These placements preserve the original RAID geometry and redundancy semantics.}
  \label{fig:data_placement}
\end{figure}
\subsection{Static RAID: Parity-Level Static Placement}
\label{sec:static_raid}
\textit{Static RAID} addresses the \emph{structural} cost of parity-based RAID. As discussed in \Cref{sec:background_raid_manage_overhead}, every stripe update in parity-based RAID must preserve parity consistency, which introduces parity-related backend writes in addition to user data updates, especially under partial-stripe updates such as RMW and RCW. This write component persists even under uniform workloads and therefore represents a workload-independent source of write amplification and wear~\cite{hu-wa-analysis09}.

Under host-visible heterogeneous flash, this structural parity cost becomes placement-sensitive. If parity chunks are stored in the same lower-endurance tier, repeated parity updates impose unavoidable program cost and wear on the capacity tier. \textit{Static RAID} targets this effect by persistently placing parity chunks on the performance tier, while keeping data chunks on the capacity tier, as shown in \Cref{fig:data_placement}. In the figure, the top two rows illustrate this \textit{Static RAID} policy: parity chunks, marked as ``P'', still follow the standard rotating RAID-5 layout across drives, while only their tier placement is redirected to the performance tier.

This design changes only the tier placement of stripe components. Stripe geometry, parity rotation, and redundancy semantics remain unchanged. By redirecting parity-related writes to the higher-endurance tier, \textit{Static RAID} reduces parity-induced wear pressure on the capacity tier and isolates the workload-independent benefit of parity placement under heterogeneous flash characteristics. Because it does not react to workload variation, \textit{Static RAID} serves as a controlled baseline for evaluating adaptive stripe placement.

\subsection{Dynamic RAID: Stripe-Level Dynamic Placement}
\label{sec:dynamic_raid}
While \textit{Static RAID} mitigates the structural parity-write component, it does not address workload-driven skew. Under skewed workloads, a small subset of stripes accumulates disproportionate activity, which is further amplified by stripe-local parity updates.

This interaction motivates \textit{Dynamic RAID}, which extends \textit{Static RAID} with adaptive stripe migration. Because parity consistency binds data and parity within a stripe, relocating only part of a hot stripe to a faster tier cannot eliminate parity-related work on the remaining components. \textit{Dynamic RAID} therefore migrates hot stripes as a whole, including both data and parity chunks, so that tier placement remains aligned with parity-based RAID semantics. In contrast, chunk-granular remapping would introduce additional metadata overhead, more complex lookup and coordination logic, and fragmented parity updates across tiers.


Using the heat-based control policy described in \Cref{sec:heat_control}, \textit{Dynamic RAID} promotes or demotes whole stripes according to bounded stripe-level activity. All migration overhead is explicitly accounted for as additional flash activity in the SSD abstraction layer. In \Cref{fig:data_placement}, the bottom two rows illustrate this \textit{Dynamic RAID} policy: whole stripes are migrated between tiers. This separation allows us to attribute reductions in stripe-level skew to adaptive placement itself, rather than to changes in RAID behavior. In this way, \textit{Dynamic RAID} complements persistent parity placement with workload-aware stripe relocation, enabling DT-RAID to address both structural and adaptive sources of stripe cost.

\subsection{Stripe-Level Cost-Aware Heat Metric}
\label{sec:heat_model}
To guide tier-aware placement, DT-RAID assigns each stripe a cost-aware heat value that reflects its effective activity under parity-based RAID. Rather than predicting absolute cost, this metric identifies stripes that incur higher effective cost due to parity-induced amplification, workload skew, and tier-dependent NAND cost asymmetry.

We model stripe activity at three levels: host accesses at the RAID interface, RAID-level amplification from parity updates, and NAND-level cost. For each stripe \(i\), let \(u_i\) denote the number of updated data chunks in an \(N\)-disk RAID array with \(M\) parity chunks, where \(1 \le u_i \le N-M\). Since each update must also maintain the \(M\) corresponding parity chunks, the RAID-level write amplification of stripe \(i\), denoted by \(A_i\), is
\begin{equation}
\label{eq:wa_raid_sim}
A_i = \frac{u_i + M}{u_i}
\end{equation}
To capture stripe activity, DT-RAID maintains separate read and write counters, denoted by \(R_i\) and \(W_i\), respectively. These counters record read and write activity and provide the measurement basis for placement control. To convert them into a cost-aware signal, we weight read and write activity according to a unified NAND cost model defined with respect to the capacity tier. Let \(t_r\), \(t_p\), and \(t_e\) denote the NAND read latency, program latency, and amortized block erase latency, respectively, and let \(A_s\) denote the SSD-level write amplification factor. We then define the normalized read and write weights as
\begin{equation}
\label{eq:nand_read_factor}
w_{r,i} =
\frac{t_r}
{t_r + (t_p + t_e)\,A_s\,A_i}
\end{equation}
\begin{equation}
\label{eq:nand_write_factor}
w_{w,i} =
\frac{(t_p + t_e)\,A_s\,A_i}
{t_r + (t_p + t_e)\,A_s\,A_i}
\end{equation}
where \(w_{r,i}\) and \(w_{w,i}\) represent the normalized read-side and write-side cost contributions of stripe \(i\), respectively.

The resulting stripe heat is defined as
\begin{equation}
\label{eq:heat_def}
H_i = w_{w,i} W_i + w_{r,i} R_i
\end{equation}

This metric exposes activity skew across stripes under heterogeneous tiers. Intuitively, stripes that are updated more frequently, experience higher parity-related amplification, and bear higher write-side costs will accumulate larger \(H_i\) values. In DT-RAID, \(H_i\) serves as the control signal for identifying hot stripes and making stripe-level placement decisions.

\subsection{Heat-Based Placement Policy}
\label{sec:heat_control}
DT-RAID uses the stripe heat metric to drive lightweight online placement control. Each stripe maintains bounded read and write counters, \(R_i\) and \(W_i\), implemented as 8-bit values in the range \([0,255]\). On every read or write access to stripe \(i\), the corresponding counter is incremented by one. As in prior history-aware and low-overhead access tracking schemes~\cite{ONeil1993LRUK,Maruf2022MultiClock}, these bounded counters provide a lightweight approximation of recent stripe activity, while the cost-aware heat \(H_i\) converts that activity into a placement-relevant signal.

To prevent unbounded heat accumulation, DT-RAID applies regulation by maintaining the average stripe heat below 127. When this bound is exceeded, stripe counters are decremented in a round-robin manner until the average falls below the threshold, avoiding counter saturation while preserving relative hotness across stripes. Placement then follows a threshold-based policy with hysteresis: a stripe is promoted when its heat reaches 140 and demoted when its heat falls below 127. This separation prevents oscillatory migration around a decision boundary. All stripes are initialized in the capacity tier and are promoted only after sufficient heat accumulates.

\section{Simulator Architecture}
\label{sec:simulator_arch}
\begin{figure}[htb]
\centering
\includegraphics[width=0.8\linewidth]{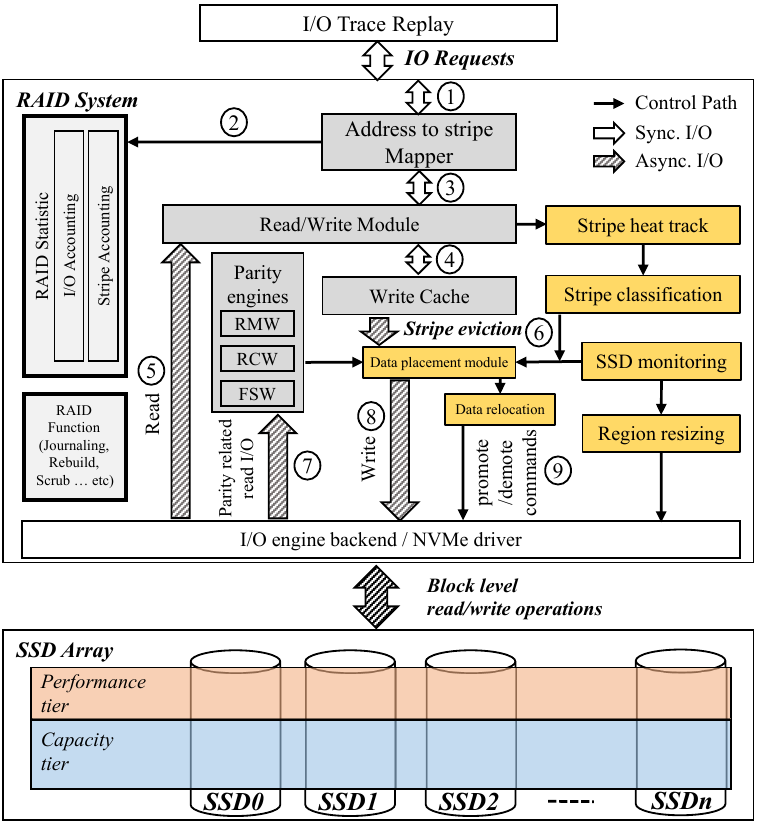}
\caption{Overall simulator architecture and request-processing flow.}
\label{fig:simulator_arch}
\end{figure}

To study DT-RAID under controlled and reproducible conditions, we implement a modular simulator that preserves parity-based RAID geometry and redundancy semantics while exposing explicit control over stripe-level tier placement. Rather than emulating full SSD internals, the framework focuses on the RAID-layer mechanisms of interest, namely parity-consistent updates, stripe-level heat tracking, tier-aware placement, and migration overhead. \Cref{fig:simulator_arch} illustrates the overall architecture and request-processing flow.

\stepnum{1}\,The simulator replays block-level I/O requests sequentially from the input trace.  
\stepnum{2}\,For each request, the address mapper translates the logical block address (LBA) into RAID-specific metadata, including the stripe identifier and chunk offset, and updates the corresponding accounting state, such as I/O statistics, stripe statistics, and stripe heat information.  
\stepnum{3}\,The translated request is then dispatched to the read/write module for further processing.  
These first three steps are common to both read and write requests and provide a uniform front-end for RAID-aware request handling.

\stepnum{4}\,For write requests, incoming data is first inserted into the write cache to model buffered write handling and absorb short-term bursts. Stripe heat is updated on each referenced access, while tier reassignment is evaluated when a stripe is selected for writeback.  
\stepnum{5}\,For read requests, the simulator first checks whether the requested data is present in the write cache. If so, the read is served directly from the cached copy. Otherwise, the framework issues a read to the SSD array and retrieves the data from the corresponding tier.  
\stepnum{6}\,When the write cache becomes full, stripes are evicted using an LRU policy.  
\stepnum{7}\,Before committing an evicted stripe, the parity engine determines the required parity update path and invokes the corresponding RMW, RCW, or FSW procedure, including any necessary parity-related reads.  
\stepnum{8}\,The resulting stripe update is then committed to the SSD array. Under \textit{Static RAID}, placement follows the fixed policy described in \Cref{sec:static_raid}. Under \textit{Dynamic RAID}, the placement module evaluates the stripe heat signal and decides whether the stripe should remain, be promoted, or be demoted. The placement module then selects the target tier.  
\stepnum{9}\,If promotion or demotion is required, the relocation module materializes the decision as explicit stripe migration between the performance and capacity tiers, and the associated movement is accounted for as additional flash activity in the SSD abstraction layer.

This unified structure allows us to evaluate parity updates, caching behavior, stripe-aware placement, and relocation overhead within a single framework. This framework also includes a control path for SSD monitoring and tier management. In principle, this path could support dynamic resizing between the performance and capacity tiers as tier occupancy changes. In this work, however, we focus on stripe placement and migration; therefore, we do not evaluate dynamic tier resizing.
\section{Evaluation}
\label{sec:evaluation}

This section evaluates the effectiveness of DT-RAID under host-visible intra-drive flash heterogeneity. Our evaluation is organized around the following hypotheses:

\begin{itemize}
\item \textbf{H1:} DT-RAID can effectively steer a significant percentage of I/O operations to the performance tier, while incurring a bounded capacity reduction.
\item \textbf{H2:} DT-RAID can improve normalized I/O acceleration under the flash-time model. 
\item \textbf{H3:} DT-RAID can improve device lifespan under endurance asymmetry across tiers.
\item \textbf{H4:} The benefits of DT-RAID remain robust under different RAID geometries, particularly different numbers of drives.
\item \textbf{H5:} The benefits of DT-RAID increase as the parity-related I/O amplification grows (RAID-5 vs RAID-6).
\item \textbf{H6:} The benefits of DT-RAID increase as the performance asymmetry between the tiers grows (SLC/TLC vs. SLC/QLC).
\end{itemize}

We first describe the experimental methodology and workloads; the remaining subsections then evaluate the validity of each hypothesis.

\begin{table}[t!]
\centering
\caption{Default DT-RAID geometry used in evaluation.}
\label{tab:raid_geometry}
\scriptsize
\begin{tabular}{l l}
\hline
Parameter & Values \\
\hline
RAID level & 5 (default) / 6 \\
Number of drives ($N$) & 4 / 8 (default) / 16 \\
Number of parity chunks ($M$) & RAID-5: 1, RAID-6: 2 \\
Chunk size & 4\,KB \\ 
Stripe size & $(N-M)\times$ chunk size \\
Parity placement & Left-symmetric manner \\
Write Amplification (SSD) & 3 \\
\hline
\end{tabular}
\end{table}
\subsection{Experimental Methodology}
\label{sec:exp-method}
We evaluate DT-RAID using the RAID-level simulation framework described in \Cref{sec:dt_raid_design,sec:simulator_arch}. The framework preserves identical RAID geometry, parity rotation, and redundancy semantics across all evaluated schemes, while exposing explicit control over stripe-level tier placement. This setup enables controlled comparison of placement policies under the same RAID organization and flash assumptions. We compare the following three configurations:
\begin{itemize}
    \item \textbf{ST RAID}: a conventional RAID design in which all data is placed in the capacity tier.
    \item \textbf{DT-S RAID}: a static placement policy that persistently places parity-related chunks on the performance tier, as described in \Cref{sec:static_raid}.
    \item \textbf{DT-D RAID}: a dynamic placement policy that performs stripe-level heat-driven migration, as described in \Cref{sec:dynamic_raid}.
\end{itemize}
Because all three configurations preserve the same stripe geometry and parity behavior, any observed differences can be attributed to tier-placement policy rather than changes in RAID semantics.

\textbf{RAID configuration.}
We model a RAID array with configurable geometry. Each stripe spans \(N\) drives and consists of \(M\) rotating parity chunks and \(N{-}M\) data chunks using a left-symmetric layout. Unless otherwise specified, we use the default geometry in \Cref{tab:raid_geometry}: an 8-disk array with 4\,KB chunks. This setting matches common OS/page I/O granularity and isolates the effect of tier placement from mapping-granularity-induced changes in FSW, RCW, and RMW behavior. To focus on RAID-level I/O behavior without interference from request-merging mechanisms, we disable I/O coalescing in this study.

\textbf{Tier and SSD model.}
Each SSD exposes two host-visible tiers with distinct latency and endurance characteristics, representing a performance-optimized tier (e.g., SLC) and a capacity-optimized tier (e.g., TLC or QLC). Each tier is parameterized by NAND read, program, and erase latencies, together with its endurance budget in program/erase cycles. The SSD abstraction accounts for per-tier flash activity by maintaining read and program counters and accumulating flash operation time. We fix the SSD write amplification factor to 3, following prior observations that report an average WAF of 3 across cloud storage workloads, and use it as a simplifying constant rather than a device-specific claim~\cite{Haas2025SSDiq}.

\begin{table}[t!] \caption{Workload analysis.}
\centering
\scriptsize
\label{tab:datasets}
\begin{tabular}{l c c c c c c}
\hline
      & IOs(M) & W(\%) & R(\%) & W (GB) & R (GB) & Total address  \\
Name  &        &       &       &        &        & space (GB)\\ \hline
hm\_0   & $4$    & $65$ & $35$ & $20.5$  & $10$    & 14.2\\
proj\_0 & $3.7$  & $88$ & $12$ & $144.3$ & $9$    & 16.6\\
prxy\_0 & $12.1$ & $97$ &  $3$ & $53.8$  & $3$    & 21.2\\
prxy\_1 & $58.1$ & $35$ & $65$ & $724.8$ & $1294$ & 69.4\\
\hline
\end{tabular}
\end{table}
\textbf{Workloads.}
We replay representative enterprise workloads from the SNIA IOTTA trace suite~\cite{msr_iotta}. \Cref{tab:datasets} summarizes their read/write ratios, total I/O counts, and working-set sizes. Each trace is replayed in its original block-level request order, and the total address space is inferred from the highest accessed LBA. Identical trace segments are used across all evaluated policies to ensure fair comparison of stripe activity, tier placement behavior, and flash cost.

\textbf{Metrics.}
Our evaluation focuses on metrics that directly support the hypotheses studied in this section. Specifically, we measure:
\begin{itemize}
    \item \textbf{Tier steering effectiveness}, including per-tier flash activity distribution and performance-tier capacity reduction.
    \item \textbf{Performance-related flash cost}, captured through accumulated flash operation time and the resulting normalized I/O acceleration.
    \item \textbf{Wear and lifetime impact}, including per-tier program activity and normalized lifespan improvement under different endurance ratios.
    \item \textbf{Sensitivity to system parameters}, including RAID geometry, parity overhead, and tier asymmetry.
\end{itemize}
These metrics allow us to evaluate not only whether DT-RAID redirects activity to the performance tier, but also whether such redirection reduces effective flash cost and improves long-term endurance under skewed workloads.

\begin{figure}[t!]
  \centering
  \includegraphics[width=1\linewidth]{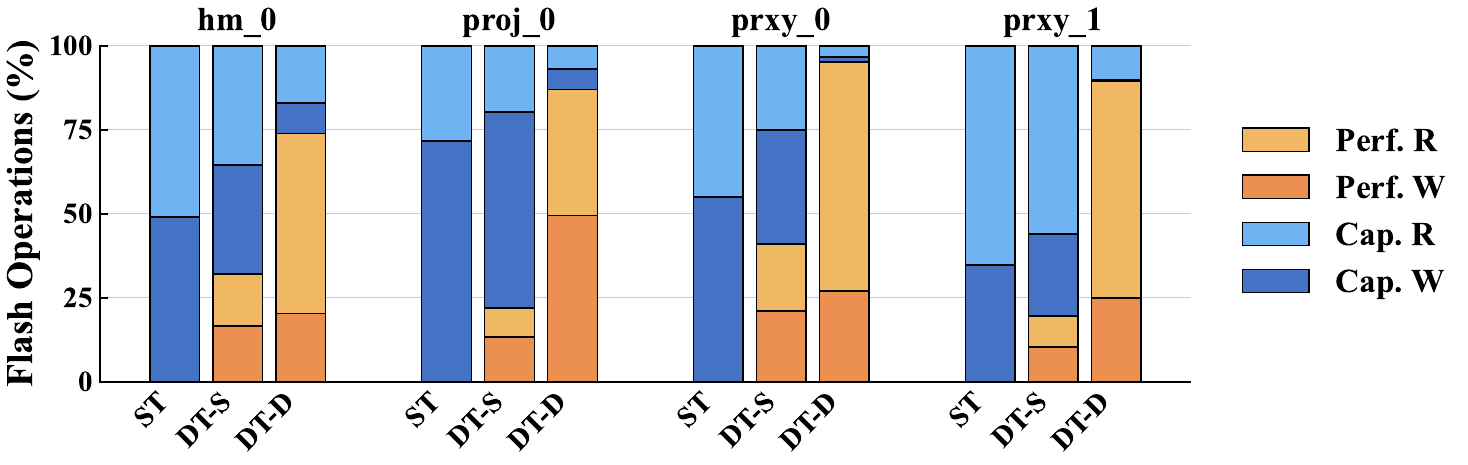}\\
  \caption{Flash operation breakdown by type and tier across workloads. Perf.\ R/W denotes performance tier read/write operations, and Cap.\ R/W denotes capacity tier read/write operations.}
  \label{fig:flash_breakdown_8disks_raid5}
\end{figure}

\subsection*{\textbf{H1:} DT-RAID Tiering Effectiveness}
\label{sec:eval_vb}

H1 examines whether DT-RAID can effectively steer RAID-induced flash activity to the performance tier while quantifying the required fast-tier footprint. We therefore evaluate DT-RAID along two dimensions: (1) how much flash activity is redirected to the performance tier, and (2) what capacity reduction is required to support such placement.
\textbf{Per-tier I/O breakdown.}
\Cref{fig:flash_breakdown_8disks_raid5} shows the tier-level breakdown of flash operations for ST, DT-S, and DT-D. Each bar is normalized to 100\% of the total flash operations observed for the corresponding workload and policy. Because all policies replay the same workload under the same RAID geometry, the key difference lies in how flash activity is distributed across tiers, with DT-D additionally accounting for migration-related flash operations. Under ST, all flash operations are directed to the capacity tier.

Across the evaluated workloads, DT-S maps between $19.6\%$ and $41.1\%$ of total NAND flash operations to the performance tier, including $10.4\%$--$21.1\%$ of writes and $8.7\%$--$20\%$ of reads. These results show that parity-only placement can shift a fraction of flash activity away from the capacity tier, even without changing RAID geometry or redundancy semantics.

DT-D further increases performance tier utilization by relocating both hot data and parity to the performance tier. Across the same workloads, DT-D maps between $74\%$ and $95.1\%$ of total NAND flash operations to the performance tier, including $20.3\%$--$49.6\%$ of writes and $37.5\%$--$68.1\%$ of reads. Compared with DT-S, dynamic placement substantially strengthens tier separation, steering most flash activity toward the performance tier while preserving the same RAID behavior.

\begin{figure}[t!]
    \centering    \includegraphics[width=0.9\linewidth]{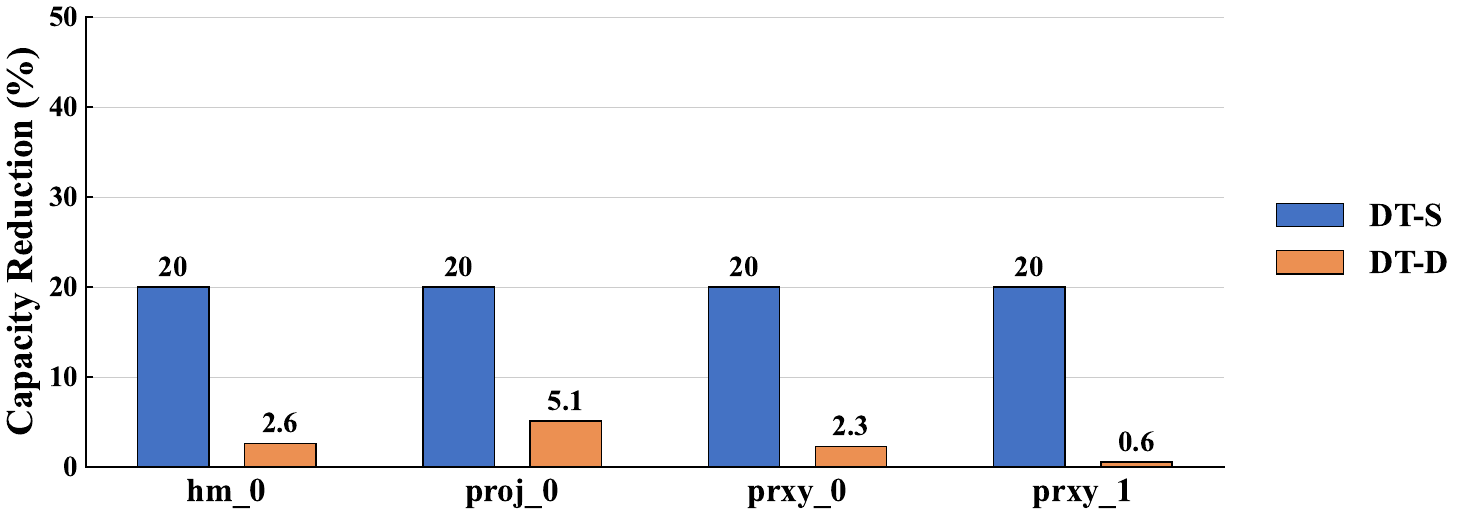}
    \caption{DT-RAID capacity reduction induced by storing parity chunks or hot stripes in the performance tier relative to a single-tier TLC-only baseline.}
    \label{fig:flash_cap_overhead}
\end{figure}

\textbf{Capacity reduction.}
Since chunks placed in the performance tier consume more raw flash capacity than equivalent chunks stored in the capacity tier, we quantify the resulting capacity reduction relative to an all-TLC baseline. \Cref{fig:flash_cap_overhead} shows the capacity reduction of DT-RAID relative to an all-TLC baseline. Let $\rho$ denote the raw-capacity ratio of multi-bit versus single-bit flash (e.g., $\rho = 3$ for TLC/SLC). Let $C_{Perf}$ and $C_{Cap}$ denote the number of chunks resident in the performance tier and capacity tier, respectively. Since each chunk placed in the performance tier consumes $\rho$ times the raw capacity of a chunk in the capacity tier, we define capacity reduction as

\begin{equation}
\mathrm{Capacity\ reduction} = (\rho - 1)\frac{C_{Perf}}{C_{Cap} + C_{Perf} \times \rho}
\end{equation}
Here, capacity reduction denotes the fraction of usable capacity lost due to mapping chunks to the performance tier.


For \textit{Static RAID}, the parity chunks are permanently placed in the performance tier, leading to a fixed capacity reduction of $20\%$ for the default 8-disk RAID-5 SLC/TLC configuration. In contrast, for \textit{Dynamic RAID}, the reduction is determined by the peak number of stripes mapped to the performance tier during the replay of each trace. Across the evaluated workloads, DT-D incurs only $0.6\%$--$5.1\%$ capacity reduction. Figures~\ref{fig:flash_breakdown_8disks_raid5} and \ref{fig:flash_cap_overhead} highlight the advantage of a dynamic tiering approach, which achieves higher performance-tier hit ratios with lower capacity reduction than a static parity-only placement policy.

\subsection*{\textbf{H2:} Normalized I/O Acceleration}
\label{sec:eval_io_accel}
Having shown that DT-RAID can redirect a large fraction of I/O activity to the performance tier, we next examine whether this steering improves overall performance. Because our framework isolates RAID-layer effects rather than modeling full device or system behavior, we estimate the reduction in total NAND operation time as a proxy for the potential performance gains of DT-RAID. A shorter I/O completion time can reduce application-level I/O wait time and simultaneously increase the SSD's internal I/O bandwidth available for flash management activities such as garbage collection, which is a known source of performance variability and latency tails~\cite{Boukhobza2025HostSideFlash,Kim2025D2FS}.

\begin{table}[t!]
\centering
\caption{Tier parameters used in the SSD abstraction.}
\label{tab:sim_tier-specs}
\begin{tabular}{l c c c}
\hline
NAND latencies & SLC & TLC & QLC\\
\hline
Page read ($L_r$) & 20 $\mu$s & 45 $\mu$s & 120 $\mu$s\\
Page program ($L_p$) & 50 $\mu$s & 400 $\mu$s & 2000 $\mu$s\\
Erase (amortized, per-page) ($L_e$) & 4 $\mu$s & 1.3 $\mu$s & 1 $\mu$s\\
\hline
\end{tabular}
\end{table}


To compute NAND operation latency, we first use the DT-RAID framework to simulate each I/O trace and collect per-tier NAND page read and program counts, as summarized earlier in \Cref{fig:flash_breakdown_8disks_raid5}. We multiply these counts by tier-specific latency values, using representative NAND latency values for each flash mode (SLC, TLC or QLC). 

\Cref{tab:sim_tier-specs} lists the latency parameters used in this study. These values are selected based on our experience with modern 3D NAND technologies and are consistent with published performance characteristics~\cite{ren2026deviceleveloptimizationtechniquessolidstate}.
Each read issued to a tier incurs the corresponding read latency \(L_r\), and each program incurs the combined program and amortized erase cost \((L_p + L_e)\). \(L_e\) denotes the amortized per-page erase latency. As bit density increases, a single erase block can correspond to more effective pages, so the erase cost is distributed over more pages, reducing the amortized erase cost per page. The total flash operation time is therefore
\begin{equation}
\small
T_{\mathrm{flash}} = \sum_{\mathrm{tiers}} \Big( N_{r,\mathrm{tier}} \cdot L_{r,\mathrm{tier}} \;+\; N_{p,\mathrm{tier}} \cdot (L_{p,\mathrm{tier}} + L_{e,\mathrm{tier}}) \Big)
\end{equation}
where \(N_{r,\mathrm{tier}}\) and \(N_{p,\mathrm{tier}}\) denote the total number of flash read and program events attributed to each tier.

Using this simplified performance model, we compare DT-RAID policies under identical RAID geometry and workload conditions and report the resulting normalized I/O acceleration. In a real deployment, additional factors such as SSD-internal buffering, controller scheduling, PCIe transfer effects, host-side software overheads, and workload-dependent parallelism would also affect end-to-end I/O performance. We leave these factors to future simulator extensions and focus here on isolating the contribution of tier-aware RAID placement itself.

\begin{figure}[t!]
    \centering
    \includegraphics[width=0.87\linewidth]{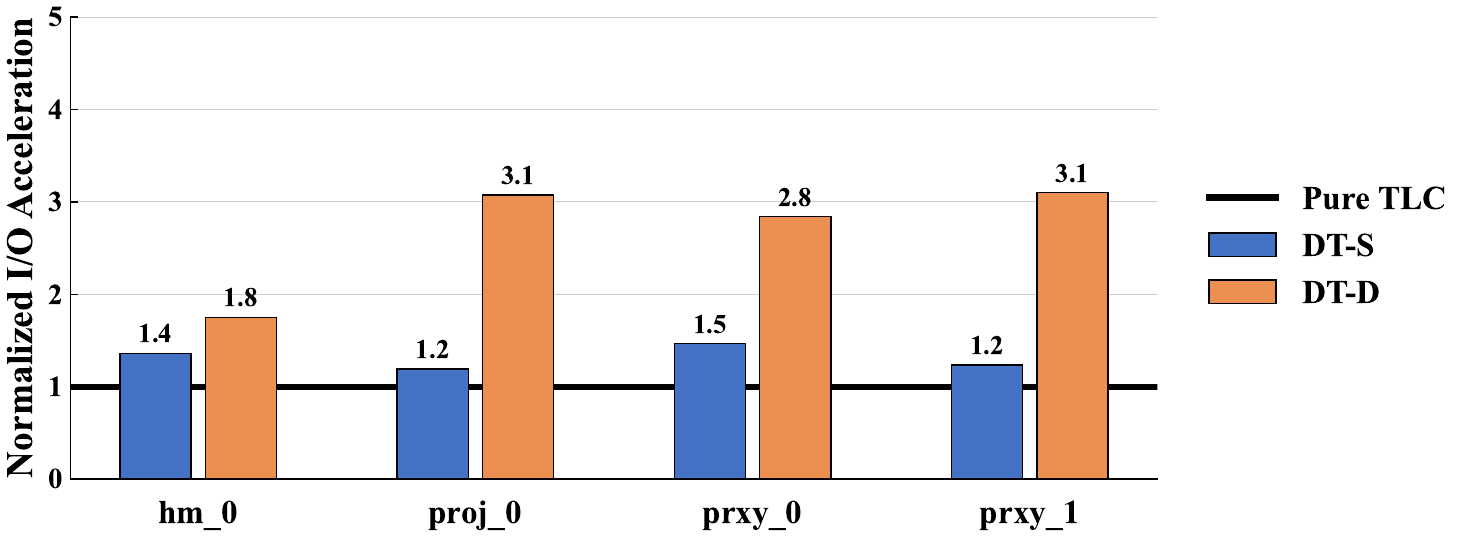}
    \caption{Normalized I/O acceleration across workloads in default RAID geometry.}
    \label{fig:io_acceleration_8disk_raid5}
\end{figure}


\Cref{fig:io_acceleration_8disk_raid5} reports the resulting normalized I/O acceleration, with ST normalized to \(1\times\). Across all workloads, DT-S consistently improves effective I/O performance over ST by steering parity-related flash activity to the performance tier. The resulting acceleration ranges from \(1.2\times\) to \(1.5\times\). These gains show that even static parity placement can reduce the effective flash service cost associated with parity maintenance.

DT-D further improves performance by relocating hot stripes to the performance tier based on the cost-aware heat metric. Under DT-D, the normalized I/O acceleration increases to \(1.8\times\) for \texttt{hm\_0}, \(3.1\times\) for \texttt{proj\_0}, \(2.8\times\) for \texttt{prxy\_0}, and \(3.1\times\) for \texttt{prxy\_1}. Compared with DT-S, these gains are substantially larger across all workloads, especially for \texttt{proj\_0}, \texttt{prxy\_0}, and \texttt{prxy\_1}. This trend is consistent with the tier-level flash breakdown in \Cref{fig:flash_breakdown_8disks_raid5}: DT-D shifts a much larger fraction of both reads and writes from the capacity tier to the performance tier. In our experiments, migration overhead is explicitly accounted for as additional read and write activity, and relocation is dominated by promotions, while demotions are rare for the evaluated workloads.

\subsection*{\textbf{H3:} Normalized Lifespan Improvement}
\label{sec:eval_lifespan_raid5}
We now analyze how DT-RAID extends array lifetime by reducing flash wear, and how this benefit depends on the endurance gap between the performance and capacity tiers.


We estimate flash wear for each I/O trace by tracking the amount of data written to each tier during simulation and converting it into flash block program/erase cycles (PECs). Because bit density determines how many blocks are needed to store a fixed amount of data, writing the same amount of data requires more SLC blocks than TLC or QLC blocks. To compare wear across heterogeneous tiers, we convert SLC PECs into multi-bit PEC equivalents using an \textit{endurance ratio} factor determined by the NAND technology and operating mode, and then compare the PECs of ST-RAID and DT-RAID to report the \emph{normalized lifespan improvement} relative to a TLC-only ST-RAID. Under the default RAID-5 configuration, we sweep the performance/capacity endurance ratio from 1:1 to 100:1 across the four workloads. A 1:1 ratio represents the worst case, where the performance tier provides no endurance advantage over the capacity tier. Ratios of 2--3:1 are achievable with minor SLC programming optimizations, whereas 4--8:1 require optimization of both program and erase operations. Much higher ratios, up to 100:1, can be achieved by permanently configuring blocks in SLC mode, trading tier-resizing flexibility for improved endurance.

\begin{figure}[t!]
    \centering
    \includegraphics[width=0.9\linewidth]{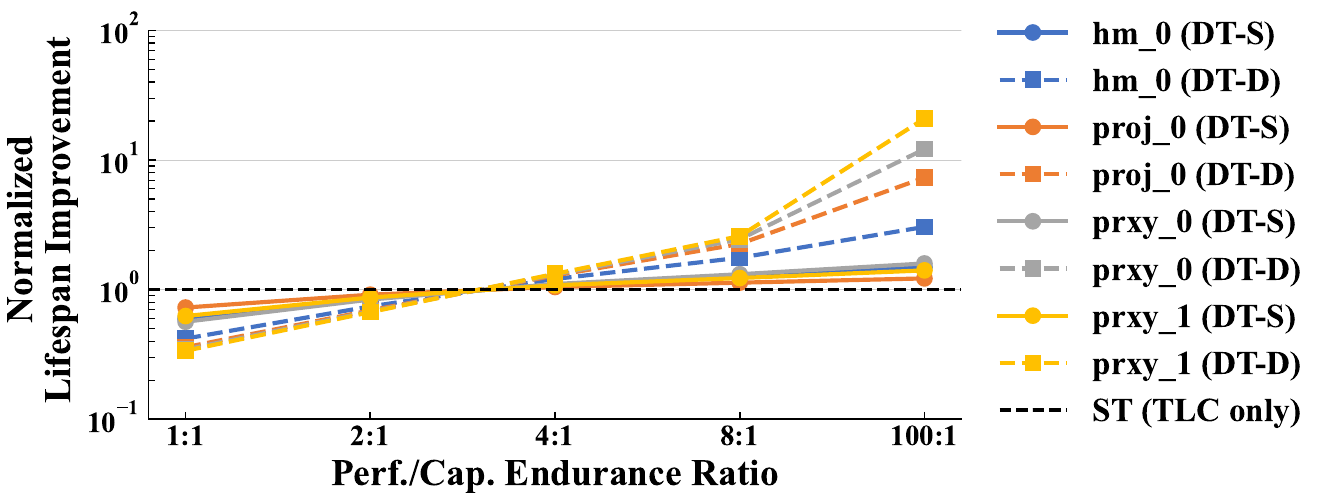}
    \caption{Normalized DT-RAID array lifespan improvement as a function of the relative performance / capacity NAND endurance ratio. All results are normalized to the ST-RAID baseline.}
    \label{fig:flash_lifespan_improvement_raid5_8disk_4k}
\end{figure}


\Cref{fig:flash_lifespan_improvement_raid5_8disk_4k} summarizes the results. 
DT-S provides limited but consistent benefits once the performance tier becomes sufficiently more durable than the capacity tier. At endurance ratios above 4:1, DT-S extends lifespan by $1.1$--$1.6\times$ versus ST-RAID. DT-D yields larger gains but also exhibits stronger workload dependence. At 4:1, DT-D extends lifespan by $1.2$--$1.3\times$; at 8:1, this increases to $1.8$--$2.6\times$; and at 100:1, it reaches $3.1$--$20.9\times$, depending on the workload skew and the fraction of writes directed to the performance tier.

Importantly, the lifespan benefits are conditional rather than universal. When the endurance ratio is below 3:1, both DT-S and DT-D can reduce lifespan, as shown in \Cref{fig:flash_lifespan_improvement_raid5_8disk_4k}. This is consistent with our wear model, which estimates lifetime consumption from endurance-normalized PECs, and with the intuition that, since writing to SLC requires $3\times$ more blocks than writing to TLC, the relative endurance ratio should be above 3:1 to compensate.

\begin{figure}[t!]
  \centering
  \includegraphics[width=1\linewidth]{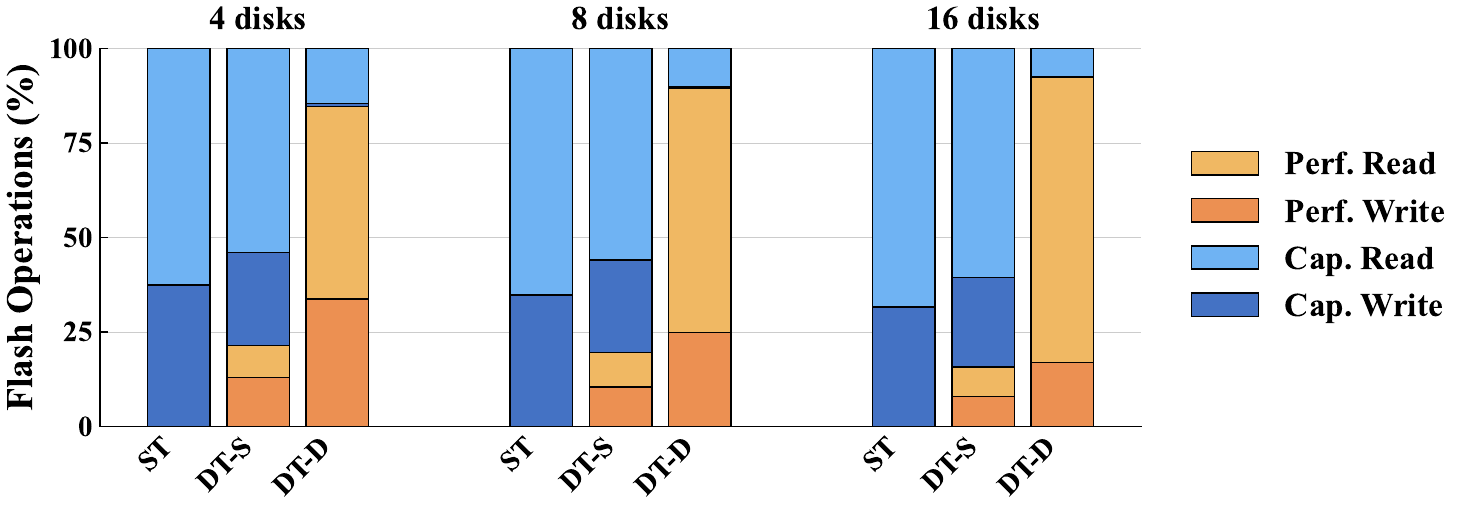}\\
  \caption{Flash operation breakdown under different RAID-5 stripe widths for \texttt{prxy\_1} with 4\,KB chunks.}
  \label{fig:riad5_ndisk_io_breakdown_prxy_1}
\end{figure}

 

\subsection*{\textbf{H4:} Sensitivity to RAID Geometry}
\label{sec:eval_stripe_width}
We next evaluate whether DT-RAID maintains its effectiveness across a range of RAID configurations. To do so, we vary the stripe width of the RAID-5 array from 4 to 16 disks while keeping the workload (\texttt{prxy\_1}) and chunk size (4 KB) constant. We then analyze how changes in stripe width influence per-tier I/O distributions as well as the capacity reduction observed in ST, DT-S, and DT-D RAID schemes.

\Cref{fig:riad5_ndisk_io_breakdown_prxy_1} reports the per-tier read/write flash I/O breakdown for $4$-, $8$- and $16$-disk RAID-5 arrays. ST I/O remains confined to the capacity tier across all configurations. For DT-S, the I/O activity directed to the performance tier decreases as the number of disks increases. In particular, the performance-tier write share drops from $24.2\%$ to $19.4\%$, while the performance-tier read share remains nearly unchanged. Correspondingly, the capacity-tier read share increases from $21.4\%$ to $28\%$. This trend is expected as DT-S is a parity-only placement policy: as stripes become wider, the parity fraction of each stripe becomes smaller, reducing the amount of flash activity that can be structurally redirected to the performance tier.


In contrast, DT-D remains effective irrespective of the stripe width. The capacity-tier write share stays extremely small, decreasing from $3.2\%$ to $0.6\%$ as the number of disks increases, while the capacity-tier read share also remains low, ranging from $3.3\%$ to $2.4\%$. 
At the same time, the share of reads handled by the performance tier rises from $48.1\%$ to $81.9\%$. This trend indicates that dynamic, stripe-level placement continues to direct the majority of flash activity to the performance tier, even as the array width increases. Overall, these results suggest that DT-D demonstrates greater robustness than DT-S as the stripe width scales.

\begin{figure}[t!]
    \centering
    \includegraphics[width=0.9\linewidth]{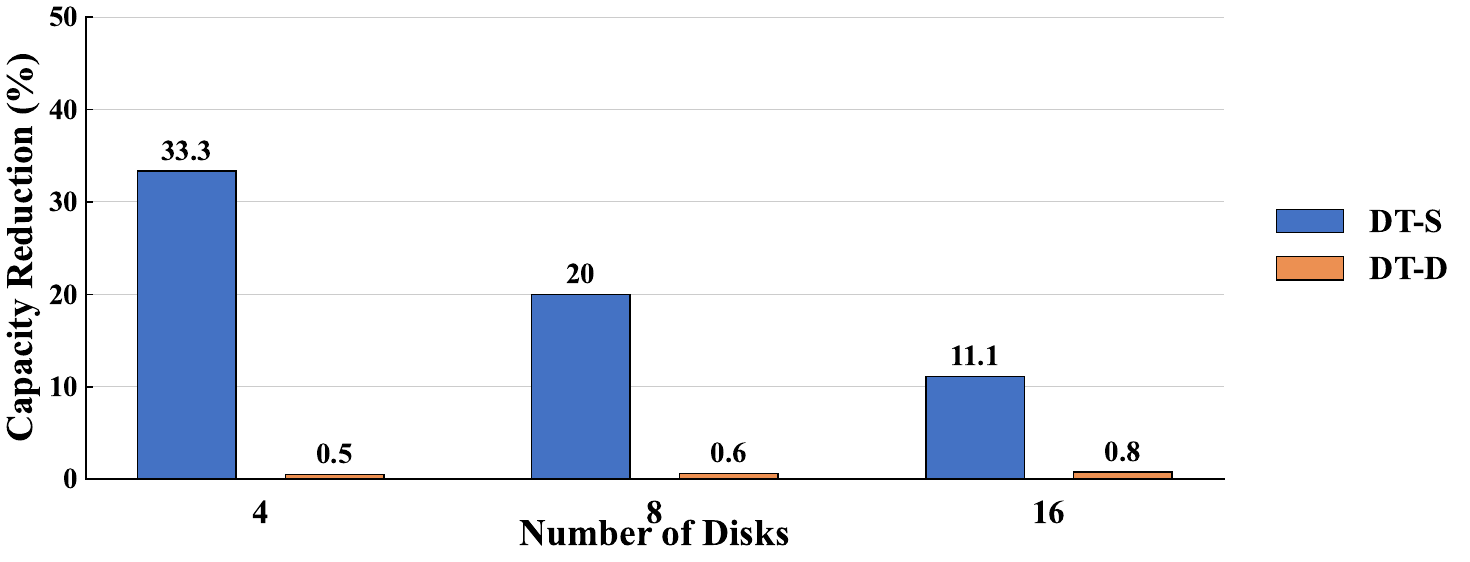}
    \caption{Capacity reduction under different RAID-5 stripe widths for \texttt{prxy\_1} with 4\,KB chunks.}
    \label{fig:riad5_ndisk_cap_overhead_prxy_1}
\end{figure}

\Cref{fig:riad5_ndisk_cap_overhead_prxy_1} reports the capacity reduction under different RAID-5 stripe widths. For DT-S, the reduction decreases from $33.3\%$ to $11.1\%$ as the number of disks increases, reflecting the smaller parity fraction of wider RAID stripes. For DT-D, the reduction remains below $1\%$ across all cases, increasing only slightly from $0.5\%$ to $0.8\%$. The results show that dynamic placement preserves effective I/O steering while introducing only minimal capacity reduction as the RAID geometry varies.

\subsection*{\textbf{H5:} Impact of Parity-related I/O Amplification}
\label{sec:eval_raid5_raid6}
\begin{figure}[t!]
    \centering
    \includegraphics[width=0.9\linewidth]{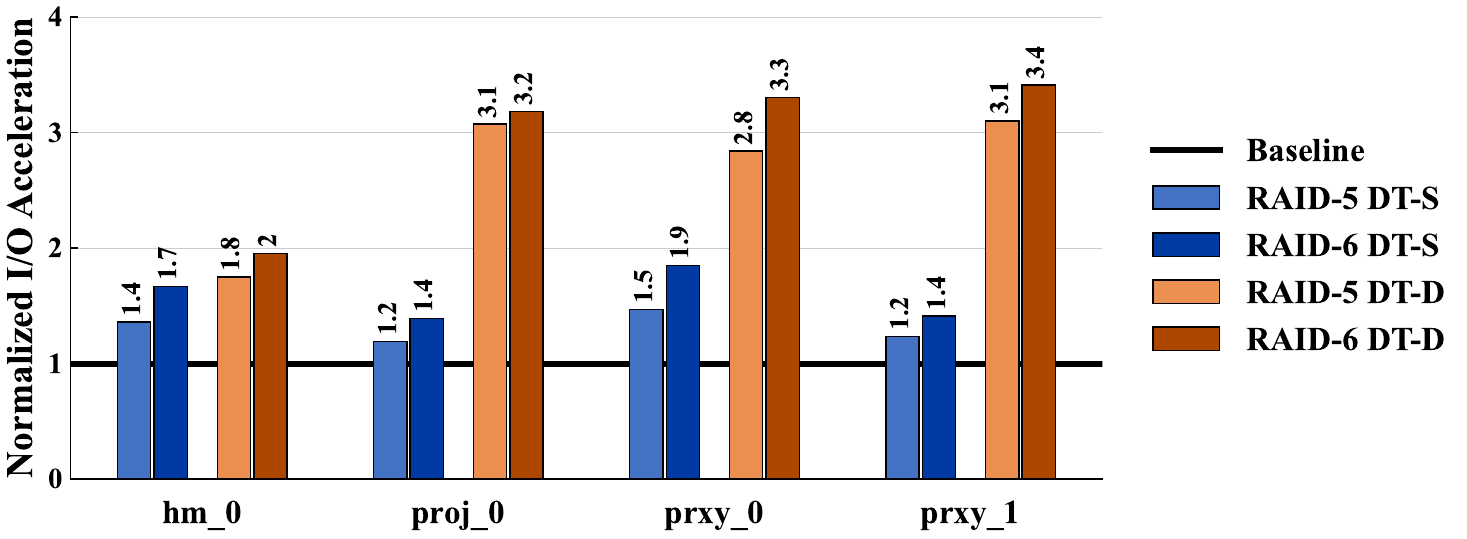}
    \caption{Normalized I/O acceleration under RAID-5 (1P+7D) and RAID-6 (2P+6D) with SLC/TLC tiers, normalized to the corresponding single-tier TLC baseline.}
    \label{fig:io_accel_raid5_raid6_8disks_4K_prxy_1}
    \vspace{-0.2cm}
\end{figure}

To evaluate how DT-RAID benefits depend on parity-related I/O amplification, we compare RAID-5 and RAID-6 under the same configuration (TLC, 8 disks, 4~KB chunk size) across the four workloads. Since RAID-6 incurs roughly twice the parity-related overhead of RAID-5, this experiment tests whether tier-aware parity placement becomes correspondingly more beneficial. \Cref{fig:io_accel_raid5_raid6_8disks_4K_prxy_1} reports the normalized I/O acceleration of ST, DT-S, and DT-D across the workloads.

DT-S shows higher acceleration under RAID-6 than RAID-5, increasing from 1.2$\times$--1.5$\times$ to 1.4$\times$--1.9$\times$ across the workloads. DT-D benefits also increase for RAID-6, with normalized I/O acceleration increasing from 1.8$\times$--3.1$\times$ under RAID-5 to 2$\times$--3.4$\times$ under RAID-6, but the relative gain is more limited. This difference is consistent with the design of the two policies. Because DT-S accelerates only parity-related activity, its benefits are directly proportional to the RAID parity overhead. In contrast, DT-D redirects a large fraction of total flash activity to the performance tier. Once the hottest stripes are placed in the performance tier, accelerating the additional parity-related I/O of RAID-6 contributes only moderate relative gains. These results indicate that higher parity I/O amplification primarily increases the relative benefits of DT-S, while DT-D remains effective irrespective of parity I/O amplification.
\subsection*{\textbf{H6:} Sensitivity to Tier Performance/Capacity Asymmetry}
\label{sec:eval_tlc_vs_qlc}
\begin{figure}[t!]
    \centering
    \includegraphics[width=0.9\linewidth]{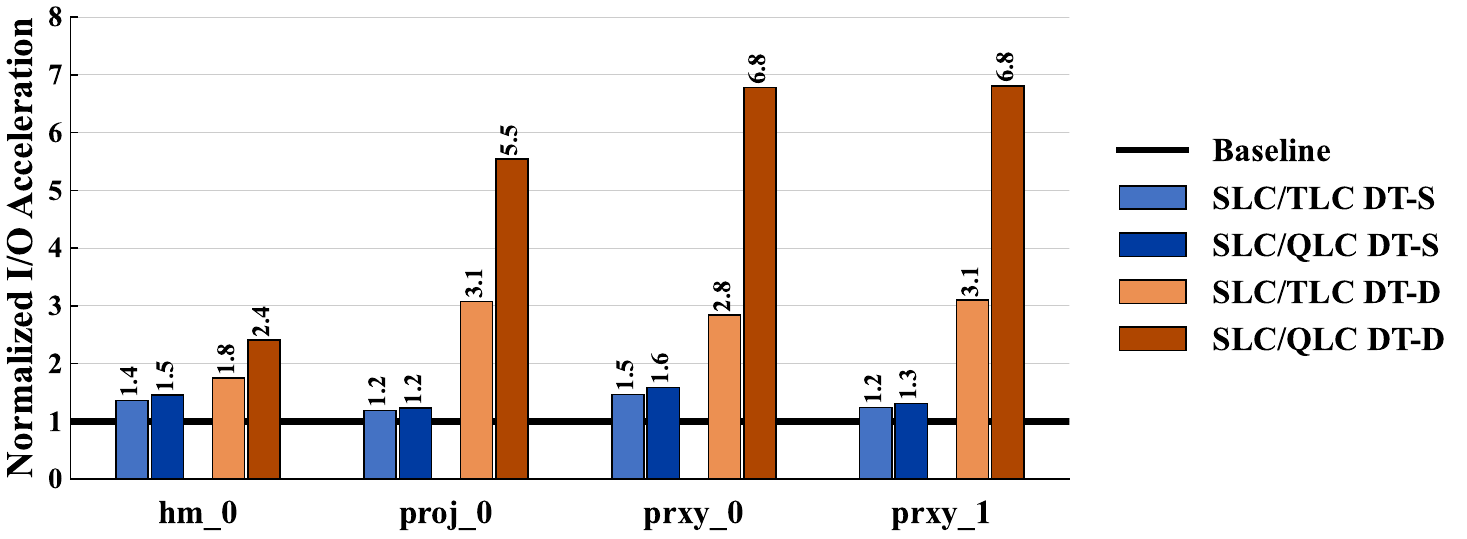}
    \caption{Normalized I/O acceleration under different tier combinations (SLC/TLC and SLC/QLC), normalized to corresponding single-tier baseline.}
    \label{fig:io_accel_qlc_tlc_raid5_8disks_4K_prxy_1}
    \vspace{-0.2cm}
\end{figure}

We further investigate DT-RAID’s behavior under varying performance disparities between the performance and capacity tiers by switching the capacity tier’s flash mode from TLC to QLC, while leaving the performance tier unchanged. This amplifies the performance asymmetry between tiers while maintaining the same RAID-5 configuration (8 disks, 4 KB chunks).
\Cref{fig:io_accel_qlc_tlc_raid5_8disks_4K_prxy_1} reports the normalized I/O acceleration under the two flash settings.

DT-S delivers moderate and relatively consistent acceleration across both SLC/TLC and SLC/QLC configurations, with normalized I/O acceleration in the 1.2$\times$–1.6$\times$ range. The results indicate that parity-only placement benefits from increased tier asymmetry. However, the overall gains are limited, as DT-S redirects only parity-related flash activity to the SLC tier.

In contrast, DT-D exhibits substantially larger improvements and a clearer dependence on tier asymmetry. Across workloads, the normalized I/O acceleration ranges from 1.8$\times$ to 3.1$\times$ with SLC/TLC and from 2.4$\times$ to 6.8$\times$ with SLC/QLC. The stronger gains for the SLC/QLC configuration are consistent with the lower baseline performance of the capacity tier. As the performance gap between the performance and capacity tiers widens, shifting I/O activity to the faster tier yields a larger reduction of the total flash operation time. 

\subsection{Discussion}
\label{sec:eval_discussion}


Overall, the evaluation supports the DT-RAID thesis of leveraging host-visible SSD tiers. For H1 and H2, DT-RAID redirects a large fraction of I/O activity from the capacity tier to the performance tier with bounded capacity reduction. DT-S provides a predictable but limited benefit by reducing structural parity I/O costs, whereas DT-D delivers larger gains by additionally steering user I/O to the performance tier for hot stripes.
For H3, the endurance benefit of DT-RAID is conditional rather than universal. Lifetime improves only when the performance tier is sufficiently more durable than the capacity tier. Otherwise, redirected write activity can make the performance tier the limiting wear factor. For H4 and H5, the results confirm that DT-S is primarily driven by parity structure: its benefits decrease as the parity fraction of wider stripes shrinks, and increase when parity-related I/O amplification grows, as in RAID-6. In contrast, DT-D remains effective across stripe widths and RAID levels as it addresses both workload hotspots and parity I/O amplification. For H6, a larger performance gap between tiers further increases the value of tier-aware placement, with DT-D benefiting the most as it redirects a larger fraction of I/O activity to the fast tier.


\section{Conclusion}
\label{sec:conclusion}
Modern SSDs already incorporate internal tiering mechanisms, typically using faster SLC flash as a buffer in front of denser, slower media. Emerging trends in industry are now moving toward exposing these internal tiers directly to the host, giving rise to host-visible heterogeneous SSDs that present multiple storage tiers with distinct performance and endurance characteristics. However, integrating such multi-tier devices within the existing storage software stack, while avoiding substantial architectural disruption, remains a challenge.

In this paper, we introduce DT-RAID, a RAID design that leverages host-visible heterogeneous tiers in SSD arrays by transparently steering I/O activity toward the performance tier. We propose two placement schemes. DT-S exploits RAID semantics to implement a straightforward static mapping that places parity chunks on the faster, higher-endurance tier. In contrast, DT-D employs dynamic placement, monitoring workload hotspots and promoting hot stripes to the performance tier. Both approaches preserve existing RAID geometry and redundancy semantics, enabling effective management of heterogeneous tiers without requiring changes to applications.

Overall, our findings position the RAID layer as a practical integration point for heterogeneous SSDs. Across a range of workloads and configurations, DT-RAID consistently steers I/O activity away from the capacity tier to the performance tier, improving modeled I/O performance and normalized lifespan while incurring a bounded and manageable space overhead. Our results indicate that the stripe granularity is a practical unit for data placement when exploiting intra-drive heterogeneity. 
\bibliographystyle{IEEEtran}
\bibliography{references-base}

@String{Computing = "Computing" }

@String{Computer = "{IEEE} Computer" }

@ARTICLE{Dong2024CloudNativeDatabases,
  author={Dong, Haowen and Zhang, Chao and Li, Guoliang and Zhang, Huanchen},
  journal={IEEE Transactions on Knowledge and Data Engineering}, 
  title={Cloud-Native Databases: A Survey}, 
  year={2024},
  doi={10.1109/TKDE.2024.3397508}}

@inproceedings{kang2014multistreamssd,
  author={Jeong‑Uk Kang and Jeeseok Hyun and Hyunjoo Maeng and Sangyeun Cho},
  title={The Multi‑streamed Solid‑State Drive},
  booktitle={6th USENIX Workshop on Hot Topics in Storage and File Systems (HotStorage)},
  year={2014}
}

@inproceedings{BjorlingBBD13,
  author    = {Matias Bj{\o}rling and Philippe Bonnet and Luc Bouganim and Niv Dayan},
  title     = {The Necessary Death of the Block Device Interface},
  booktitle = {Proceedings of the 6th Biennial Conference on Innovative Data Systems Research (CIDR 2013)},
  year      = {2013},
}

@inproceedings{2025KVSSD_park,
author = {Park, Chanyoung and Lee, Jungho and Liu, Chun-Yi and Kang, Kyungtae and Kandemir, Mahmut Taylan and Choi, Wonil},
title = {AnyKey: A Key-Value SSD for All Workload Types},
year = {2025},
isbn = {9798400706981},
doi = {10.1145/3669940.3707279},
booktitle = {Proceedings of the 30th ACM International Conference on Architectural Support for Programming Languages and Operating Systems},
}

@inproceedings{10.1145/3297663.3310302,
author = {Talluri, Sacheendra and \L{}uszczak, Alicja and Abad, Cristina L. and Iosup, Alexandru},
title = {Characterization of a Big Data Storage Workload in the Cloud},
year = {2019},
isbn = {9781450362399},
doi = {10.1145/3297663.3310302},
booktitle = {Proceedings of the 2019 ACM/SPEC International Conference on Performance Engineering},
}

@article{li2020cloudblock,
author = {Li, Jinhong and Wang, Qiuping and Lee, Patrick P. C. and Shi, Chao},
title = {An In-depth Comparative Analysis of Cloud Block Storage Workloads: Findings and Implications},
year = {2023},
address = {New York, NY, USA},
issn = {1553-3077},
doi = {10.1145/3572779},
journal = {ACM Transactions on Storage}
}

@article{tang2023cloudonomics,
      title={Rethinking the Cloudonomics of Efficient I/O for Data-Intensive Analytics Applications}, 
      author={Chunxu Tang and Yi Wang and Bin Fan and Beinan Wang and Shouwei Chen and Ziyue Qiu and Chen Liang and Jing Zhao and Yu Zhu and Mingmin Chen and Zhongting Hu},
      year={2023},
      eprint={2311.00156},
      journal={arXiv},
      primaryClass={cs.DC},
}

@article{Boukhobza2025HostSideFlash,
author = {Boukhobza, Jalil and Olivier, Pierre and Lim, Wen Sheng and Chen, Liang-Chi and Hsieh, Yun-Shan and Wu, Shin-Ting and Ho, Chien-Chung and Huang, Po-Chun and Chang, Yuan-Hao},
title = {A Survey on Flash-Memory Storage Systems: A Host-Side Perspective},
year = {2025},
issue_date = {August 2025},
publisher = {Association for Computing Machinery},
address = {New York, NY, USA},
issn = {1553-3077},
doi = {10.1145/3723167},
journal = {ACM Transactions on Storage}
}

@INPROCEEDINGS{Radu2019HybridFlashCTRL,
  author={Stoica, Radu and Pletka, Roman and Ioannou, Nikolas and Papandreou, Nikolaos and Tomic, Sasa and Pozidis, Haris},
  booktitle={2019 IEEE 27th International Symposium on Modeling, Analysis, and Simulation of Computer and Telecommunication Systems (MASCOTS)}, 
  title={Understanding the Design Trade-Offs of Hybrid Flash Controllers}, 
  year={2019},
  doi={10.1109/MASCOTS.2019.00025}
}

@inproceedings{Kim2025D2FS,
author = {Kim, Juwon and Lee, Seungjae and Oh, Joontaek and Shin, Dongkun and Won, Youjip},
title = {{D2FS}: device-driven filesystem garbage collection},
year = {2025},
isbn = {978-1-939133-45-8},
booktitle = {Proceedings of the 23rd USENIX Conference on File and Storage Technologies},
}

@inproceedings{Cho2024AERO,
author = {Cho, Sungjun and Kim, Beomjun and Cho, Hyunuk and Seo, Gyeongseob and Mutlu, Onur and Kim, Myungsuk and Park, Jisung},
title = {{AERO}: Adaptive Erase Operation for Improving Lifetime and Performance of Modern NAND Flash-Based SSDs},
year = {2024},
isbn = {9798400703867},
doi = {10.1145/3620666.3651341},
booktitle = {Proceedings of the 29th ACM International Conference on Architectural Support for Programming Languages and Operating Systems},
}

@article{Kim2025REO,
  author  = {Beomjun Kim and Myungsuk Kim},
  title   = {REO: Revisiting Erase Operation for Improving Lifetime and Performance of Modern NAND Flash-Based SSDs},
  journal = {Electronics},
  year    = {2025},
  doi     = {10.3390/electronics14040738}
}

@misc{2025MixedModeSSD,
  title        = {Mixed Mode {SSD}},
  author       = {Klemm, Mike},
  year         = {2025},
  howpublished = {Future Memory \& Storage (FMS) 2025 Proceedings, Session SSDT-203-1},
  organization = {KIOXIA Inc.},
}

@inproceedings{Yoo2020Reinforcement,
author = {Yoo, Sangjin and Shin, Dongkun},
title = {Reinforcement learning-based {SLC} cache technique for enhancing {SSD} write performance},
year = {2020},
booktitle = {Proceedings of the 12th USENIX Conference on Hot Topics in Storage and File Systems},
}

@INPROCEEDINGS{Zgang2019SPASSD,
  author={Zhang, Wenhui and Cao, Qiang and Jiang, Hong and Yao, Jie and Dong, Yuanyuan and Yang, Puyuan},
  booktitle={2019 IEEE 37th International Conference on Computer Design (ICCD)}, 
  title={{SPA-SSD}: Exploit Heterogeneity and Parallelism of 3D {SLC}-{TLC} Hybrid {SSD} to Improve Write Performance}, 
  year={2019},
  doi={10.1109/ICCD46524.2019.00088}}

@ARTICLE{Cai2017ErrorCharacters,
  author={Cai, Yu and Ghose, Saugata and Haratsch, Erich F. and Luo, Yixin and Mutlu, Onur},
  journal={Proceedings of the IEEE}, 
  title={Error Characterization, Mitigation, and Recovery in Flash-Memory-Based Solid-State Drives}, 
  year={2017},
  doi={10.1109/JPROC.2017.2713127}}

@article{Haas2025SSDiq,
author = {Haas, Gabriel and Lee, Bohyun and Bonnet, Philippe and Leis, Viktor},
title = {{SSD-iq}: Uncovering the Hidden Side of {SSD} Performance},
year = {2025},
doi = {10.14778/3749646.3749694},
journal = {Proceedings of the VLDB Endowment}
}

@inproceedings{Shu2023dRAID,
  author    = {Jin Shu and Zhen Zhu and Jinyu Wang and Yang Hu and Jidong Zhai and Youyou Lu},
  title     = {Disaggregated {RAID} Storage in Modern Datacenters},
  booktitle = {Proceedings of the 28th ACM International Conference on Architectural Support for Programming Languages and Operating Systems (ASPLOS)},
  year      = {2023},
  doi       = {10.1145/3582016.3582027}
}

@inproceedings{2022scalaraid,
author = {Yi, Shushu and Yang, Yanning and Tang, Yunxiao and Zhou, Zixuan and Li, Junzhe and Yue, Chen and Jung, Myoungsoo and Zhang, Jie},
title = {{ScalaRAID}: optimizing linux software {RAID} system for next-generation storage},
year = {2022},
isbn = {9781450393997},
doi = {10.1145/3538643.3539740},
booktitle = {Proceedings of the 14th ACM Workshop on Hot Topics in Storage and File Systems},
}

@article{2025scalaafa-journal,
author = {Zhang, Jie and Yi, Shushu and Pan, Xiurui and Xu, Yiming and Li, Qiao and Li, Qiang and Wang, Chenxi and Mao, Bo and Jung, Myoungsoo},
title = {Enhancing the Performance of Next-Generation {SSD} Arrays: A Holistic Approach},
year = {2025},
address = {New York, NY, USA},
issn = {1553-3077},
doi = {10.1145/3736588},
journal = {ACM Transactions on Storage},
}

@article{2024straid-journal,
author = {Wang, Shucheng and Cao, Qiang and Jiang, Hong and Lu, Ziyi and Yao, Jie and Chen, Yuxing and Pan, Anqun},
title = {Explorations and Exploitation for Parity-based {RAIDs} with Ultra-fast {SSDs}},
year = {2024},
issue_date = {February 2024},
publisher = {Association for Computing Machinery},
address = {New York, NY, USA},
journal = {ACM Transactions on Storage},
issn = {1553-3077},
doi = {10.1145/3627992}
}

@ARTICLE{2024hybraid,
  author={Karimi, Maryam and Salkhordeh, Reza and Brinkmann, André and Asadi, Hossein},
  journal={IEEE Transactions on Parallel and Distributed Systems}, 
  title={{HybRAID}: A High-Performance Hybrid {RAID} Storage Architecture for Write-Intensive Applications in All-Flash Storage Systems}, 
  year={2024},
  doi={10.1109/TPDS.2024.3429336}}

@inproceedings{2024asymetricraid,
author = {Jiao, Ziyang and Kim, Bryan S.},
title = {Asymmetric {RAID}: Rethinking {RAID} for {SSD} Heterogeneity},
year = {2024},
isbn = {9798400706301},
doi = {10.1145/3655038.3665952},
booktitle = {Proceedings of the 16th ACM Workshop on Hot Topics in Storage and File Systems},
}

@inproceedings{berg2020cachelib,
author = {Berg, Benjamin and Berger, Daniel S. and McAllister, Sara and Grosof, Isaac and Gunasekar, Sathya and Lu, Jimmy and Uhlar, Michael and Carrig, Jim and Beckmann, Nathan and Harchol-Balter, Mor and Ganger, Gregory R.},
title = {The {CacheLib} caching engine: design and experiences at scale},
year = {2020},
isbn = {978-1-939133-19-9},
booktitle = {Proceedings of the 14th USENIX Conference on Operating Systems Design and Implementation},
}

@inproceedings {ZhichaoCao2020RocksDBworkloads,
author = {Zhichao Cao and Siying Dong and Sagar Vemuri and David H.C. Du},
title = {Characterizing, Modeling, and Benchmarking {RocksDB} {Key-Value} Workloads at Facebook},
booktitle = {18th USENIX Conference on File and Storage Technologies (FAST 20)},
year = {2020},
isbn = {978-1-939133-12-0},
}

@misc{msr_iotta,
  author       = {{MSR}},
  title        = {{MSR Cambridge Traces (SNIA IOTTA Trace Set)}},
  howpublished = {{SNIA IOTTA Trace Repository}, Geo Kuenning (Ed.), Storage Networking Industry Association},
  year         = {2007},
}

@article{yadgar2021ssd,
  title   = {{SSD}-based workload characteristics and their performance implications},
  author  = {Yadgar, Gala and Gabel, Moshe and Jaffer, Shehbaz and Schroeder, Bianca},
  journal = {ACM Transactions on Storage},
  year    = {2021}
}

@article{yang2016write,
author = {Yang, Yue and Zhu, Jianwen},
title = {Write Skew and {Zipf} Distribution: Evidence and Implications},
year = {2016},
issue_date = {August 2016},
publisher = {Association for Computing Machinery},
address = {New York, NY, USA},
issn = {1553-3077},
doi = {10.1145/2908557},
journal = {ACM Transactions on Storage},
}

@inproceedings{wu2025skewness,
author = {Wu, Haonan and Xu, Erci and Wang, Ligang and Hong, Yuandong and Niu, Changsheng and Shi, Bo and Zhu, Lingjun and He, Jinnian and Wu, Dong and Zhang, Weidong and Wang, Qiuping and Wang, Changhong and Chen, Xinqi and Xue, Guangtao and Chen, Yi-Chao and Ding, Dian},
title = {Hey Hey, My My, Skewness Is Here to Stay: Challenges and Opportunities in Cloud Block Store Traffic},
year = {2025},
isbn = {9798400711961},
doi = {10.1145/3689031.3696068},
booktitle = {Proceedings of the Twentieth European Conference on Computer Systems},
}

@inproceedings{maneas2022operational,
  title     = {Operational characteristics of {SSDs} in enterprise storage systems: A large-scale field study},
  author    = {Maneas, Stathis and Mahdaviani, Kaveh and Emami, Timothy and Schroeder, Bianca},
  booktitle = {Proceedings of the 20th USENIX Conference on File and Storage Technologies (FAST)},
  year      = {2022},
}

@article{Xue2024LakehouseScale,
author = {Xue, Maryann and Bu, Yingyi and Somani, Abhishek and Fan, Wenchen and Liu, Ziqi and Chen, Steven and van Hovell, Herman and Samwel, Bart and Mokhtar, Mostafa and Korlapati, RK and Lam, Andy and Ma, Yunxiao and Ercegovac, Vuk and Li, Jiexing and Behm, Alexander and Li, Yuanjian and Li, Xiao and Krishnamurthy, Sriram and Shukla, Amit and Petropoulos, Michalis and Paranjpye, Sameer and Xin, Reynold and Zaharia, Matei},
title = {Adaptive and Robust Query Execution for Lakehouses at Scale},
year = {2024},
issue_date = {August 2024},
issn = {2150-8097},
doi = {10.14778/3685800.3685818},
journal = {Proceedings of the VLDB Endowment},
}

@inproceedings{Hu2024CharLLMinDatacenter,
author = {Hu, Qinghao and Ye, Zhisheng and Wang, Zerui and Wang, Guoteng and Zhang, Meng and Chen, Qiaoling and Sun, Peng and Lin, Dahua and Wang, Xiaolin and Luo, Yingwei and Wen, Yonggang and Zhang, Tianwei},
title = {Characterization of large language model development in the datacenter},
year = {2024},
isbn = {978-1-939133-39-7},
booktitle = {Proceedings of the 21st USENIX Symposium on Networked Systems Design and Implementation},
}

@manual{KIOXIAFL6Datasheet,
  author       = {{KIOXIA Corporation}},
  title        = {{KIOXIA Enterprise SSD FL6 Series Product Brief}},
  year         = {2021},
  organization = {KIOXIA Corporation},
  url          = {https://americas.kioxia.com/content/dam/kioxia/shared/business/ssd/enterprise-ssd/asset/productbrief/eSSD-FL6-product-brief.pdf}
}

@manual{SolidigmD7P5810Brief,
  author       = {{Solidigm}},
  title        = {{Solidigm D7-P5810 Product Brief}},
  year         = {2023},
  organization = {Solidigm},
  url          = {https://www.solidigm.com/content/dam/solidigm/en/site/products/data-center/d7/p5810/documents/solidigm-d7-p5810-product-brief.pdf}
}

@manual{MicronION6500,
  author       = {{Micron}},
  title        = {{Micron 6500 ION SSD Series}},
  year         = {2023},
  organization = {Micron},
  url          = {https://assets.micron.com/adobe/assets/urn:aaid:aem:55527cf3-550d-4873-b99b-4f02404ac36d/renditions/original/as/6500-ion-nvme-ssd-tech-prod-spec.pdf}
}

@manual{NVMeBaseSpec22,
  title        = {{NVM Express Base Specification, Revision 2.2}},
  year         = {2025},
  organization = {NVM Express, Inc.},
  note         = {March 11, 2025},
  url          = {https://nvmexpress.org/wp-content/uploads/NVM-Express-Base-Specification-Revision-2.2-2025.03.11-Ratified-1.pdf}
}

@article{ren2026deviceleveloptimizationtechniquessolidstate,
      title={Device-Level Optimization Techniques for Solid-State Drives: A Survey}, 
      author={Tianyu Ren and Yajuan Du and Jinhua Cui and Yina Lv and Qiao Li and Chun Jason Xue},
      year={2026},
      eprint={2507.10573},
      journal={arXiv},
      primaryClass={cs.AR},
}

@INPROCEEDINGS{Maruf2022MultiClock,
  author={Maruf, Adnan and Ghosh, Ashikee and Bhimani, Janki and Campello, Daniel and Rudoff, Andy and Rangaswami, Raju},
  booktitle={2022 IEEE International Symposium on High-Performance Computer Architecture (HPCA)}, 
  title={{MULTI-CLOCK}: Dynamic Tiering for Hybrid Memory Systems}, 
  year={2022},
  doi={10.1109/HPCA53966.2022.00072}}

@inproceedings{ONeil1993LRUK,
author = {O'Neil, Elizabeth J. and O'Neil, Patrick E. and Weikum, Gerhard},
title = {The LRU-K page replacement algorithm for database disk buffering},
year = {1993},
isbn = {0897915925},
doi = {10.1145/170035.170081},
booktitle = {Proceedings of the 1993 ACM SIGMOD International Conference on Management of Data},
}

@ARTICLE{Wu2024FIRMTree,
  author={Wu, Shin-Ting and Chen, Pin-Jung and Huang, Po-Chun and Shih, Wei-Kuan and Chang, Yuan-Hao},
  journal={IEEE Transactions on Computer-Aided Design of Integrated Circuits and Systems}, 
  title={FIRM-Tree: A Multidimensional Index Structure for Reprogrammable Flash Memory}, 
  year={2024},
  doi={10.1109/TCAD.2024.3445809}}

@article{Ren2024NearFree,
author = {Ren, Tianyu and Li, Qiao and Lv, Yina and Ye, Min and Guan, Nan and Jason Xue, Chun},
title = {{Near}-{Free} Lifetime Extension for 3D {NAND} Flash via Opportunistic Self-Healing},
year = {2024},
doi = {10.1109/TCAD.2024.3447225},
journal = {IEEE Transactions on Computer-Aided Design of Integrated Circuits and Systems},
}

@inproceedings{hu-wa-analysis09,
  title={Write amplification analysis in flash-based solid state drives},
  author={Hu, Xiao-Yu and Eleftheriou, Evangelos and Haas, Robert and Iliadis, Ilias and Pletka, Roman},
  booktitle={Proceedings of SYSTOR 2009: The Israeli Experimental Systems Conference},
  pages={10},
  year={2009},
  organization={ACM}
}
\end{document}